\documentclass[useAMS,usenatbib,a4paper]{mn2e}
\usepackage{graphicx}
\usepackage{aas_macros}
\usepackage{multirow}
\usepackage{hyperref}
 
\title[Mid-infrared spectroscopy of M31]{Mid-infrared spectroscopy of the Andromeda galaxy}
\author[D. Hemachandra et al.]
{D. Hemachandra$^{1}$,
P. Barmby$^{1}$\thanks{E-mail: pbarmby@uwo.ca}, 
E. Peeters$^{1,2}$\thanks{E-mail: epeeters@uwo.ca},  
S.P. Willner$^{3}$, 
M.L.N. Ashby$^{3}$,
H.A. Smith$^{3}$, 
\newauthor 
K.D. Gordon$^{4}$,
D.A. Smith$^{4}$,
and
G.G. Fazio$^{3}$\\
$^{1}$Department of Physics and Astronomy, University of Western Ontario, London, ON, N6A 3K7, Canada\\
$^{2}$SETI Institute, 189 Bernardo Avenue, Suite 100, Mountain View, CA 94043, USA\\
$^{3}$Harvard-Smithsonian Center for Astrophysics, Cambridge, MA 02138, USA\\
$^{4}$Space Telescope Science Institute, 3700 San Martin Drive, Baltimore, MD 21218, USA
}

\begin{document}

\date{}

\maketitle

\label{firstpage}

\begin{abstract}
We present {\sl Spitzer}/Infrared Spectrograph (IRS) 5--21~$\mu$m spectroscopic maps towards 12 regions in the Andromeda galaxy (M31). 
These regions include the nucleus, bulge, an active region in the star-forming ring, and 9 other regions chosen to cover a range of mid-to-far-infrared colours. 
In line with previous results, PAH feature ratios (6.2~$\mu$m and 7.7~$\mu$m features compared to the 11.2~$\mu$m feature) measured from our extracted M31 spectra, except the nucleus, strongly correlate. The equivalent widths of the main PAH 
features, as a function of metallicity and radiation hardness, are consistent with those observed for other nearby spiral and starburst galaxies. 
Reprocessed data from the ISOCAM instrument on the {\it Infrared Space Observatory} agree with the IRS data; early reports of suppressed 6--8~\micron\ features and enhanced 11.3~$\mu$m feature intensity and FWHM apparently resulted from background-subtraction problems.
The nucleus does not show any PAH emission but does show strong silicate emission at 9.7~$\mu$m. Furthermore, different spectral features (11.3~$\mu$m PAH emission, silicate emission and [NeIII] 15.5~$\mu$m line emission) have distinct spatial distributions in the nuclear region: the silicate emission is strongest towards the stellar nucleus, while the PAH emission peaks 15\arcsec\ north of the nucleus. The PAH feature ratios at this position are atypical with strong emission at 11.2~\micron\ and 15--20~\micron\ but weak emission at 6--8~\micron. The nucleus itself is dominated by stellar light giving rise to a strong blue continuum and silicate emission. 
\end{abstract}


\section{Introduction}
Mid-infrared spectra provide a unique diagnostic tool to understand the physical conditions in the interstellar medium of galaxies. 
The rich range of spectral features (including Polycyclic Aromatic Hydrocarbons (PAHs), atomic fine structure lines (e.g. Ne, S) and the
amorphous silicate feature centred at 9.7~$\mu$m provide information on dust properties, radiation field and star formation. 
With the advent of infrared space telescopes, such as the Infrared Space Observatory (ISO, \citealt{Kessler1996}) and 
the {\em Spitzer} Space Telescope \citep{spitzer2004},  mid-infrared spectroscopy has become an important method
of investigating the infrared emission from galaxies. 

PAHs are known as the carrier of the ubiquitous mid-infrared emission bands (e.g. \citealt{Allamandola1989}, \citealt{puget89}).
They are large hydrocarbon molecules consisting of $\sim$50--100 carbon atoms \citep{Tielens2008}. 
The main PAH features are seen at 3.3, 6.2, 7.7, 8.6, 11.2 and 12.7~$\mu $m (e.g., \citealt{Gillett:73}, \citealt{Geballe:85}, \citealt{Peeters:toledo:11}). 
These bands are attributed to the vibrational de-excitation of PAH molecules through bending and stretching modes of C-H and C-C bonds (e.g. \citealt{Allamandola1989}, \citealt{puget89}). 
The 6 to 8 micron features are thought to originate mostly from ionized PAHs and the 3.3 and 11.2~$\mu$m 
emission bands from neutral PAHs (e.g. \citealt{Hudgins:rev:04} and reference therein, \citealt{Hony:oops:01},  \citealt{Galliano2008}). 

The relative strength of the PAH features vary spatially within extended objects and from galaxy-to-galaxy \citep[e.g.][]{Galliano2008}. While extinction does influence the individual PAH bands to different degrees \citep[e.g.][]{Brandl2006, Stock:13}, the observed spread in PAH intensity ratios is dominated by the PAH charge balance \citep[e.g.][]{Galliano2008}. In addition, PAH intensity ratios change more drastically close to active galactic nuclei where the overall strength of PAH emission also gets weaker (\citealt{Roche1991}, \citealt{Smith:2007lr}). \citet{Smith:2007lr} found that the mid-infrared 
spectra from some weak AGNs show suppressed 6 to 8~$\mu$m PAH features (by up to a factor of 10 in strength) but are bright at 11.2~$\mu$m. One possible explanation for this behaviour is that AGNs alter the grain composition by selective destruction of small PAHs and/or excite the PAHs. Alternatively, the PAH emission is modified by the low star formation intensity in the centres of many AGNs \citep{Smith:2007lr}. 

Previous studies of nearby galaxies indicate that metallicity and radiation hardness can both affect the PAH emission (e.g. \citealt{Madden:00}, \citealt{Beirao:06}, \citealt{Engelbracht_2008}, \citealt{Munoz:09}). In a sample of nearby starburst galaxies \citet{Engelbracht_2008} found the PAH equivalent widths (EQWs) to decrease 
with increasing radiation hardness; however, \citet{Brandl2006} found no such correlation within their starburst sample. In addition, PAH EQWs show an anti-correlation with metallicity in star-forming galaxies. This variation of PAHs among galaxies has also been observed within H~{\sc ii} regions 
of M101 \citep{Gordon:2008lr}. But there are no other investigations done on a single star-forming galaxy with sufficiently high resolution to see whether the correlations mentioned above hold within a galaxy similar to the Milky Way.

The amorphous silicate feature at 9.7~$\mu$m is another aspect of the mid-infrared spectra of galaxies and in particular their nuclei.  \citet{Spoon2007} 
classified infrared galaxies based on the equivalent width of the 6.2~$\mu$m PAH feature and the strength of the 9.7~$\mu$m silicate emission or absorption feature. 
They  found galaxies spread along two distinct branches: one in which silicate absorption strength was anti-correlated with PAH
equivalent width, and another in which the weak silicate feature strength did not depend on the 6.2~$\mu$m equivalent width.
Silicate {\em emission} at 9.7~$\mu$m has also been observed  towards several galaxies including LINERS, Seyfert galaxies, ULIRGs, and QSOs \citep[e.g.][]{Sturm2005, Hao2005, Spoon2007, Smith2010, Mason2012} 
and can be used to constrain the geometry and structure of the emitting nuclear region \citep{Mason2009}.

M31 with its proximity \citep[$785\pm25$ kpc; ][]{Mcc2005} and rich observational databases provides the most detailed view of a star forming galaxy similar 
to the Milky Way. The active star forming ring visible in 8~$\mu$m  {\em Spitzer}/IRAC images \citep{Barmby2006lr} provides evidence of abundant PAHs in M31. 
However, ISOCAM spectro-imaging observations \citep{1998Cesarsky} of four regions in M31 including the nucleus and bulge 
of this galaxy showed very odd PAH spectra, bright at 11.2~$\mu$m with weak or no  6.2, 7.7, and 8.6~$\mu$m bands. 
Investigating this unusual PAH emission was the main motivation for the work described in this paper.

The centre of M31 has a complicated physical structure. It hosts a very inactive supermassive black hole with a mass of 
$0.7-1.4 \times 10^8$~M$_{\sun}$ \citep{Bacon2001, Bender2005} and also has a lopsided nuclear disk  with two stellar 
components \citep{Lauer1993} and an A-star cluster \citep{Bender2005}. While M31's nucleus is known to be inactive from an 
X--ray perspective \citep{Li2011}, mid-infrared indicators of its nuclear activity, such
as infrared excess or spectral features of silicates,  have received relatively little attention. 
The higher spatial resolution available in observations of  a very nearby galaxy like M31, compared to 
luminous, distant objects such as ultra-luminous infrared galaxies \citep{Spoon2007} or nearby Seyferts \citep{Mason2009},
makes exploring its mid-infrared spectrum worthwhile.

We employed mid-infrared spectral maps from the {\em Spitzer}/Infrared Spectrograph (IRS) from 12 regions of M31 for a further investigation of 
its infrared properties. This sample includes the nucleus, bulge, an active region in the star-forming ring (all previously observed by ISOCAM), and 9 
other regions chosen to cover a range of properties as described in Section~\ref{sect:irs_obs}. 
We also obtained the highly processed version of ISOCAM observations of M31 and compare them with the IRS results in Section~\ref{sect:iso_vs_irs}. 
Section~\ref{sect:pah_ratios} discusses PAH intensity ratios.
In Section~\ref{sect:eqw_rh}, we investigate the relationship between PAH equivalent widths, radiation 
hardness, and metallicity and compare to that found by \citet{Engelbracht_2008} and \citet{Gordon:2008lr}.  
Section~\ref{sect:nucleus} discusses the dust properties of the nucleus. The paper is summarized in Section~\ref{sect:summary}.

\section{Observations and data reduction}

\subsection{IRS observations}
\label{sect:irs_obs}

\begin{figure}
\centering
\includegraphics[width = 4.5 cm,angle=270]{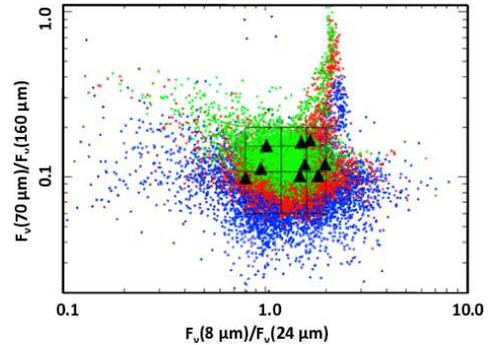}
\caption{$8 - 24/70 - 160$ $\mu$m colour-colour diagram of M31 obtained from IRAC and MIPS imaging. Colour-coding of points represents
24~$\mu$m flux, from faintest (blue, dark grey) to average (green, light grey) to brightest (red, medium grey). The plot is divided into 9 regions (black grid), and the observations were made to 
cover those regions subject to a 24~$\mu$m brightness cut. The black triangles indicate colours of the regions we observed.}
\label{colourmaps}
\end{figure}

\begin{figure*}
\centering
\includegraphics[scale=0.9]{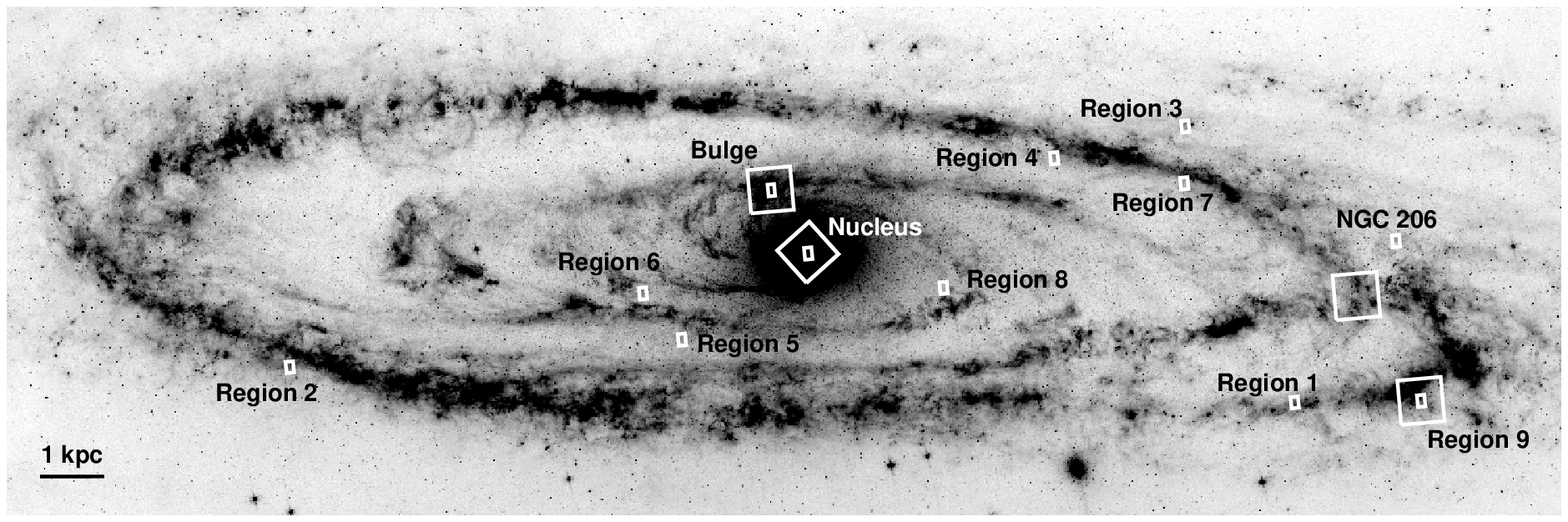}
\caption{An 8 micron negative IRAC image of M31 \citep[dark colours indicate higher flux,][]{Barmby2006lr}. Small white rectangles ($30\arcsec\times50\arcsec$) show the regions that we observed, and larger squares ($192\arcsec\times192\arcsec$) show the regions observed by  \citet{1998Cesarsky}.
\label{m31}
}
\end{figure*}

We obtained mid-infrared spectral maps of 12 regions in M31 using the {\em Spitzer}/IRS instrument \citep{IRS2004} covering wavelengths from 5 to 21 microns. 
A background observation was also made off the galaxy 
along the minor axis for use in background subtraction from the data cubes.
The 12 target regions include the nucleus, the `bulge' and `active' regions previously observed by ISOCAM (the latter is Region 9 in our sample), 
and 9 other regions selected to cover a range of metallicities and dust temperatures.%
\footnote{One additional spectral map in M31 is available in the {\it Spitzer} archive (AOR key 12019200);
unfortunately it does not cover the 5--13~$\mu$m region, which is the major focus of this paper. We therefore elected to not include these observations.} 
These 9 regions were chosen by convolving the IRAC 8~$\mu$m \citep{Barmby2006lr}
and MIPS \citep{gordon06a} maps to the same resolution and constructing an $8 - 24/70 - 160$ $\mu$m colour-colour diagram (Figure~\ref{colourmaps}).
This colour space was used to give a rough definition of the types of spectral energy distribution; the 
dense region in the plot was split into a 3x3 grid and one pixel in each grid region (subject to a 24$\mu$m brightness cut)
was selected for spectroscopy. 
The locations of the observed regions are shown in Figure~\ref{m31}, and 
their coordinates and metallicities are given in Table~\ref{regions}.  
Except for regions 5 and 8, all of our mapped regions contain an  H~{\sc ii} region with
an optical spectroscopic metallicity measurement by \citet{Sanders_2011}; Table ~\ref{regions} gives the measurements from the method 
\citet{Sanders_2011} denote ``N06 N2''  \citep{Nagao2006}. For regions 5 and 8 we estimated metallicities using the M31 radial gradient
that  \citet{Sanders_2011} fit to the H{\sc ii} region ``N06 N2'' measurements: $12+\log{\rm [O/H]} = 9.09 - 0.00195 R_{\rm gc}$.

\begin{table*}
 \centering
 \begin{minipage}{90mm}
\caption{Spitzer/IRS Target Locations in M31
\label{regions}}
\begin{tabular}{lccrl}
\hline Name & R.A. (J2000) & Decl. (J2000) & ${R_{\rm gc}}^b$ & $12+\log({\rm O/H})^c$
\\
 \hline
Nucleus$^a$ & $00^{\rmn{h}}42^{\rmn{m}}44\fs31$ & $41\degr16\arcmin09\farcs4$  & 0.0 & \\
Bulge$^a$   & $00^{\rmn{h}}42^{\rmn{m}}35\fs00$ & $41\degr21\arcmin01\farcs0$  & 4.7 &$8.90\pm0.03$\\
Region 1    & $00^{\rmn{h}}41^{\rmn{m}}30\fs41$ & $40\degr43\arcmin07\farcs8$  & 12.4 &$9.20\pm0.20$\\
Region 2    & $00^{\rmn{h}}45^{\rmn{m}}22\fs85$ & $41\degr38\arcmin53\farcs1$  & 13.0 &$9.07\pm0.02$\\
Region 3    & $00^{\rmn{h}}40^{\rmn{m}}37\fs37$ & $41\degr01\arcmin29\farcs4$  & 12.1 &$8.85\pm0.01$\\
Region 4    & $00^{\rmn{h}}41^{\rmn{m}}17\fs86$ & $41\degr07\arcmin09\farcs8$  & 8.7 &$8.89\pm0.06$\\
Region 5    & $00^{\rmn{h}}43^{\rmn{m}}39\fs57$ & $41\degr19\arcmin03\farcs1$  & 7.0 &$8.93\pm0.08$\\
Region 6    & $00^{\rmn{h}}43^{\rmn{m}}35\fs72$ & $41\degr23\arcmin15\farcs0$  & 4.3 &$8.73\pm0.08$\\
Region 7    & $00^{\rmn{h}}40^{\rmn{m}}53\fs98$ & $40\degr58\arcmin58\farcs9$  & 8.7 &$8.40\pm0.08$\\
Region 8    & $00^{\rmn{h}}42^{\rmn{m}}21\fs60$ & $41\degr07\arcmin17\farcs4$  & 3.1 &$8.94\pm0.08$\\
Region 9$^a$& $00^{\rmn{h}}41^{\rmn{m}}00\fs00$ & $40\degr36\arcmin20\farcs3$  & 13.5 &$8.86\pm0.02$\\
NGC~206     & $00^{\rmn{h}}40^{\rmn{m}}20\fs20$ & $40\degr44\arcmin54\farcs0$  & 9.8 & \\
Background  & $00^{\rmn{h}}44^{\rmn{m}}41\fs80 $ & $40\degr58\arcmin56\farcs0$  & 29.5 & \\
\hline
\end{tabular}
{$^a$Regions with ISOCAM data.\\
$^b$De-projected galactocentric distance in kpc.\\ 
$^c$Metallicities from \citet{Sanders_2011}, except for Regions 5 and 8 where metallicities are estimated from the radial metallicity profile.
}
\end{minipage}
\end{table*}

\begin{figure}
\centering
\includegraphics[scale=0.27]{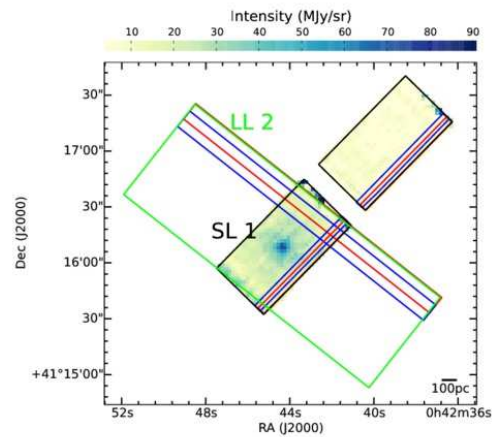}
\caption{The 7.6~$\mu$m plane constructed from the SL1 data cube of the nucleus, showing the arrangement of slits used to cover the region. 
Black boxes outline the footprint of the SL1 maps (the off-centre SL1 map is from observations made
when SL2 was centred) and the green box outlines the LL2 map. 
Blue and red slits show how  each map was covered using overlapping slit positions.
\label{slits}
}
\end{figure}

For our observations we used the IRS Short-Low (SL) and Long-Low (LL) modules, which cover wavelengths from 5 to 21 microns. 
These modules have resolving power in the range 60--130. Each low-resolution module is divided into two sub-slits 
which provide spectroscopy in either first or second order. They are denoted as SL1 (7.5--14.5~$\mu$m), SL2 (5.2--7.6~$\mu$m),
LL1 (20.5--38.5~$\mu$m, not used in these observations), and LL2 (14.5--20.75~$\mu$m). The data were obtained in `no peak-up' mode resulting in a positional uncertainty of 1\arcsec. 
All M31 regions were observed in September 2007 as part of G. G. Fazio's Guaranteed Time (program ID 40032). 
The map size was based on the size 
of the IRS slits (SL: $3.6\arcsec \times 57\arcsec$, LL: $10.5\arcsec \times 168\arcsec$). Each region was covered by 18 overlapping observations 
of the SL slit and 11 overlapping observations of the LL slit making the map size $32\arcsec \times 57\arcsec$ for SL and $58\arcsec \times 168\arcsec$ for LL. 
Figure~\ref{slits} shows an example of the slit arrangement. For the brighter regions (nucleus, bulge), ramp times of 14 s (SL) and 30 s (LL) were used, 
while for the fainter regions, ramp times of 60 and 120 s were used respectively. Background observations were taken with each module (2 per ramp time). 
Because all of the targets are in the same part of the sky, a common background observation was used for multiple targets to subtract the background emission. 

\subsection{IRS Data Reduction}
\label{sect:irs_data}

The data were reduced through the SSC pipeline (ver. S17.2.0), and the maps were assembled using the CUBISM program \citep{Smith:2007fk}. 
Bad pixel removal was also done using CUBISM, and the background observations were used to subtract the background emission from these cubes 
following the method outlined by \citet{Gordon:2008lr}. Spectra were extracted using a $30\arcsec\times50\arcsec$   rectangular aperture,
which corresponds to $114\times190$~pc at the distance of M31.
The aperture size was selected to cover the overlapping area of the SL and LL modes; all the IRS maps cover more area than  considered here.
In order to study the spatial variation of the emission near the nucleus, we also extracted spectra from two smaller regions
within that map; these will be further discussed in Section~\ref{sect:nucleus}.
The spectrum of NGC 206 is very noisy and was removed from our analysis. 

There is wavelength overlap between the SL1 and SL2 spectra and also between the SL1 and LL2 spectra.
To generate a single spectrum for each M31 region it is necessary to combine the spectra and
account for photometric offsets between them. Such offsets are commonly seen in IRS spectra extracted
from extended regions. 
For most regions, the SL1 and SL2 flux densities were
quite well-matched over the wavelength overlap region ($7.5 < \lambda< 7.6\mu$m),
and those two orders were joined without offset.
In two cases with offsets between SL1 and SL2, SL2 and the `bonus order' SL3 were well-matched.
We therefore combined the  SL1 and SL3 
spectra by 
adding a constant  to the SL1 spectra so that they matched the SL3 average. The SL1 and SL2 orders
were then combined as described above.
After this procedure there was still a noticeable mis-match between the SL and LL spectra. We addressed this
by scaling the SL spectra to match IRAC 8~$\mu$m fluxes \citep{Barmby2006lr} in the same apertures used to extract the IRS spectra.
An extended source  aperture correction of 0.824 was applied to the IRAC fluxes; this value was computed 
using the formula in the IRAC Data Handbook \citep{SpitzerIIH} with
the radius of a circular aperture having the same area as the extraction aperture.
The {\em Spitzer} synthetic photometry software \citep{SpitzerDAC} 
was then used to quantify the colour correction for each spectrum, i.e. the
multiplicative factor $K$ between the IRAC photometry over its broad bandpass and the IRS flux
density at the centre of the bandpass.  For each spectrum we computed the additive
offset between the colour-corrected IRS flux density and the IRAC aperture-corrected flux density,
and used this to correct the SL spectra; this method was also used by \citet{Sandstrom12}. The SL spectra corrected with additive offsets were closely-matched to the LL spectra.
Although a multiplicative offset could also have been used to match the IRS photometry to IRAC,
the resulting change in spectral slope increased the size of the SL/LL offset.

\begin{figure}
\centering
\includegraphics[width = 7.4cm]{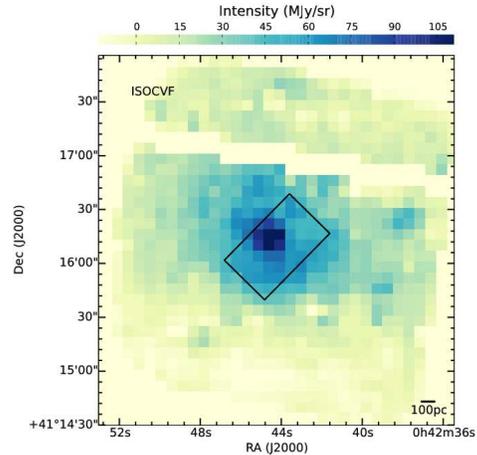}
\caption{11.2~$\mu$m negative image of the ISOCAM data cube from the nucleus of M31. 
The black box shows the size of the aperture ($30\arcsec\times50\arcsec$) used to extract IRS spectra.}
\label{isonuc}
\end{figure}

\begin{figure*}
\centering
\includegraphics[scale=0.45]{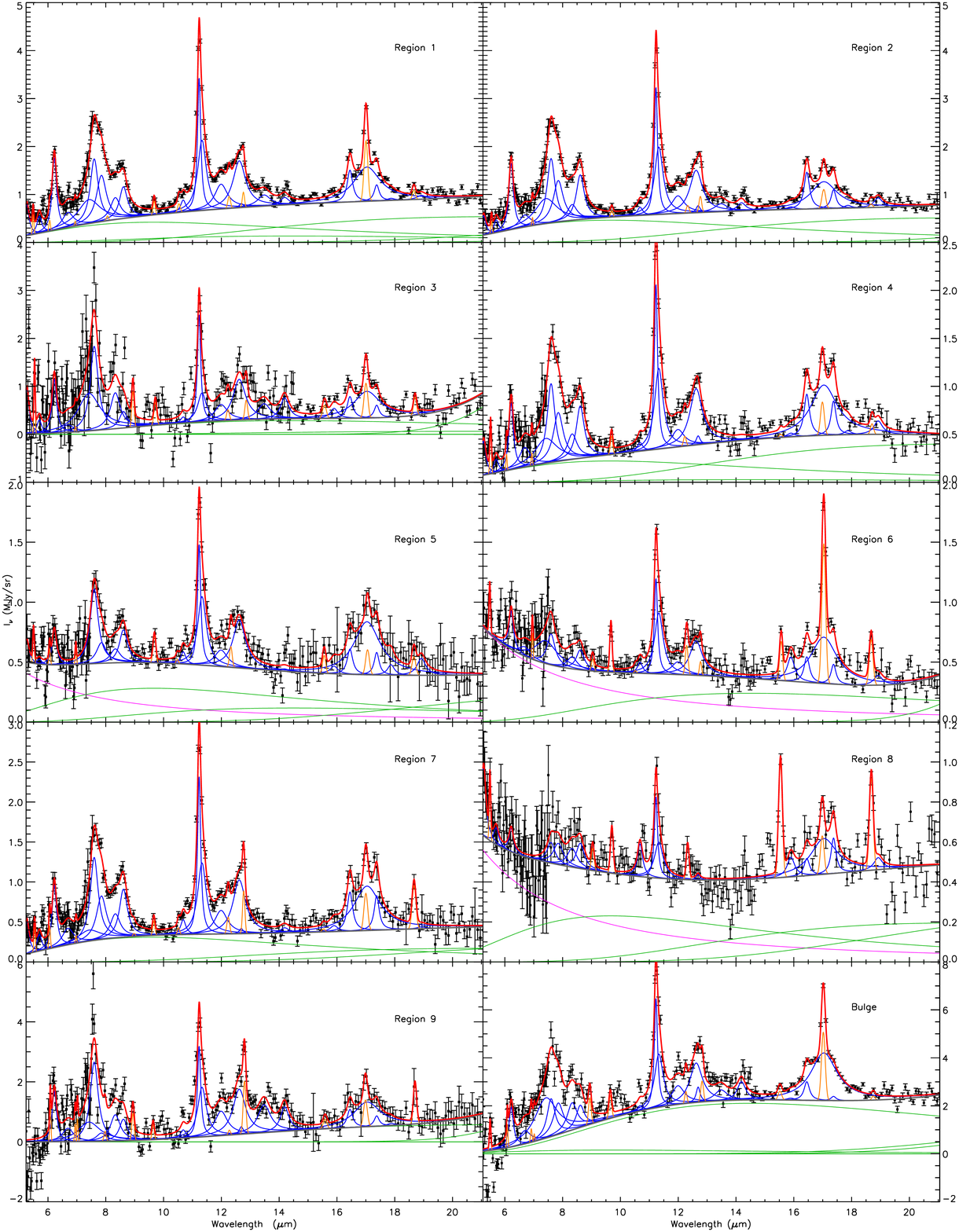}
 \caption{Observed IRS spectra and detailed PAHFIT decompositions (see Section~\ref{sect:pahfit}). Black circles show the observed data, and green, blue, orange, magenta and red lines represent the modelled
dust continua, PAH features, atomic and molecular lines, starlight continuum and the fit respectively. The grey line shows the total modelled continuum. 
Vertical scales differ in different panels. Spectra from the nucleus and NGC 206 are not shown here.
}
\label{PAHFITplots}
\end{figure*}

\subsection{ISOCAM Data Reduction}
\label{sect:iso_data}

To compare our results with those of  \citet{1998Cesarsky}, we retrieved the highly processed ISOCAM data provided by \citet{Boulanger_F_2005}  
for the nucleus, bulge, and region 9. 
The ISOCAM data were obtained with circular variable filters (CVFs) over a $3\arcmin \times 3\arcmin$ field of view at a scale of 6\arcsec\ per pixel. 
The wavelength range covered was 5.15--16.5~$\mu$m at a resolution of $\lambda/\Delta \lambda \approx 45$; the ISOCAM instrument is described by \citet{cesarsky1996}.
An example image of the ISOCAM data is shown in Figure~\ref{isonuc}.  For the three regions, we extracted spectra using the same 
$30\arcsec\times50\arcsec$ aperture as for the IRS data.

\section{Spectral characteristics and feature measurements}
\label{sect:data_analysis}

The final processed IRS spectra are shown in  Figure \ref{PAHFITplots}, except for the nucleus, discussed in Section~\ref{sect:nucleus}.
All of the main PAH features, including the 6.2, 7.7, 8.6 and 11.2~$\mu$m bands, 
are clearly visible for all the regions shown here.
The IRS spectra also show atomic line emission such as [Ar~{\sc ii}], [Ar~{\sc iii}], [S~{\sc iii}], [S~{\sc iv}], [Ne~{\sc ii}], [Ne~{\sc iii}] 
and molecular H$_{2}$ emission at 12.3~$\mu$m. Some of the spectra display a contribution to the continuum from starlight emission.

\subsection{ISOCAM versus IRS}
\label{sect:iso_vs_irs}

As mentioned in the Introduction, based on ISOCAM observations of four regions in M31, \citet{1998Cesarsky} reported 
suppression of the 6--8~$\mu$m PAH bands and a strong 11.2~$\mu$m PAH band.
The 11.2~$\mu$m band profile was found to vary from region to region.  
In addition, \citet{Pagani_1999} confirmed that the star-forming ring in M31 seemed to show very weak PAH emission in the 6 to 8~$\mu$m region. 
However, the IRS spectra presented here in Figure~\ref{PAHFITplots} do not show such unusual behaviour. 
Indeed all regions, except the nucleus, show a normal mid-infrared spectrum similar to other nearby star-forming galaxies. 

\begin{figure}
\centering
\includegraphics[scale=0.35]{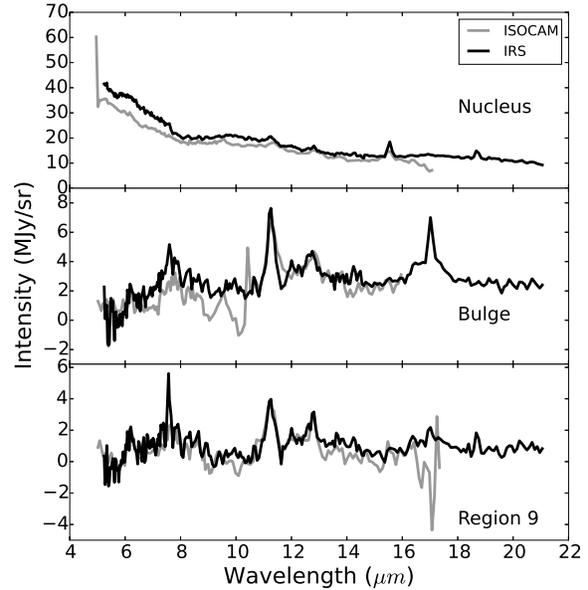}
\caption{ Comparison of  IRS and re-processed ISOCAM spectra for the Nucleus (top), Bulge (middle) and Region 9 (bottom) in M31.}
\label{ISOnIRS}
\end{figure}

Until 2005, ISOCAM-CVF data were not properly background subtracted, and they were contaminated with zodiacal emission and stray light. 
Differential spectra between regions of relatively strong and weak emission were previously used to overcome this problem 
(more details about the differential spectra are given by \citealt{1998Cesarsky}, \citealt{Pagani_1999}). 
In 2005, all ISOCAM-CVF data were reprocessed  and corrected for the zodiacal emission \citep{Boulanger_F_2005}. 
We obtained these newly-processed ISOCAM spectra from three regions in our IRS sample (see Section~\ref{sect:iso_data}) 
in order to compare them with the corresponding IRS spectra.
Figure~\ref{ISOnIRS} shows that although the relative feature intensities  in the IRS and ISOCAM 
spectra differ in detail, the spectral shapes are almost identical;
the re-processed ISOCAM data do not agree with the previous differential spectra.
Neither the bulge nor Region 9 shows any depletion in  6 to 8~$\mu$m features. The new ISOCAM reduction appears to eliminate the discrepancies between ISOCAM and IRS, resulting
in less `strange' infrared spectra for M31. For the remainder of this work, we discuss only the IRS spectra.

\subsection{PAHFIT}
\label{sect:pahfit}
The PAH features in  IRS spectra are often blended with neighbouring PAH features and atomic lines. 
Therefore measuring the strength of PAH features is difficult.  To achieve this task a tool called PAHFIT, introduced by \citet{Smith:2007lr}, was used. 
PAHFIT is an IDL  based tool designed for decomposing {\em Spitzer} IRS low-resolution spectra of PAH emission sources.
PAHFIT uses six main components to fit the surface brightness. These are starlight continuum, featureless thermal dust continuum, 
pure rotational lines of H$_2$, fine-structure lines, dust emission features and dust extinction. The starlight is represented by  blackbody 
emission at a fixed temperature of 5000~K, and the dust continuum is represented by 8 modified blackbodies (emissivity proportional to $\nu^2$)  
at fixed temperatures of 35, 40, 50, 65, 90, 135, 200, and 300~K. The final fit obtained with PAHFIT does not necessarily use
all eight dust components.
The infrared extinction is considered as a combination of a power law plus silicate absorption features with peaks at 9.7 and 18~$\mu$m. 
Line features are represented by Gaussian profiles with widths set by the instrumental resolution
and dust features are represented by Drude profiles; more details about PAHFIT are given by \citet{Smith:2007lr}.

Initial attempts at fitting the spectra with PAHFIT showed that some components were negligible, and
to avoid over-fitting we re-ran the fits without these components.
None of the IRS spectra shows significant silicate absorption around 9.7 or 18~$\mu$m, and the extinction calculated by PAHFIT 
was almost zero for all the initial fits. Except for four regions (the bulge, Region 5, Region 6 and Region 8), the starlight contribution is also negligible. This is consistent with the fact that regions 5, 6, and 8 are the innermost of the regions. 
We adjusted the PAHFIT input parameters to fix extinction to zero for all regions and starlight to zero for all but the four regions above.
Typically only two or three thermal dust components had significant power in our fits, but we did not fix the unused components to zero.
Regions 3 and 9 were found to have very low dust continuum emission compared to other spectra,
possibly because of noisy data at short wavelengths. However the other features in these spectra appear to
be fit correctly. The spectrum of the nucleus shows silicate emission, which is not included in PAHFIT;  
our modifications of PAHFIT to analyze this spectrum are discussed in Section~\ref{sect:nucleus}.

\begin{table*}
 \centering
 \begin{minipage}{200mm}
\caption{PAH Emission Line Strengths$^a$}
{\scriptsize
  \begin{tabular}{l c c  c  c  c  c  c  c  c  c c }
\hline
    {Region }&{5.7 $\mu$m  }&{6.2 $\mu$m  }&{7.7 $\mu$m$^b$  }&{8.3 $\mu$m  }&{8.6 $\mu$m  }&{10.7 $\mu$m  }&{11.2 $\mu$m$^b$  }&{12.0 $\mu$m  }&{12.7 $\mu$m$^b$  }&{17.0 $\mu$m$^b$  } \\
 \hline
       Bulge  & \dots & $1.32 \pm 0.08$ & $7.7 \pm 0.6$ & $1.1 \pm 0.2$ & $0.7 \pm 0.1$ & $0.07 \pm 0.03$ & $1.85 \pm 0.08$ & $0.49 \pm 0.05$ & $1.02 \pm 0.09$ & $1.39 \pm 0.04$\\
    Region 1  & $0.36 \pm 0.05$ & $1.20 \pm 0.04$ & $3.8 \pm 0.2$ & $0.46 \pm 0.04$ & $0.50 \pm 0.03$ & $0.08 \pm 0.01$ & $1.18 \pm 0.02$ & $0.32 \pm 0.02$ & $0.54 \pm 0.02$ & $0.58 \pm 0.02$\\
    Region 2  & $0.27 \pm 0.03$ & $1.10 \pm 0.03$ & $3.7 \pm 0.2$ & $0.33 \pm 0.04$ & $0.70 \pm 0.03$ & $0.053 \pm 0.008$ & $1.13 \pm 0.02$ & $0.23 \pm 0.01$ & $0.51 \pm 0.03$ & $0.51 \pm 0.02$\\
    Region 3  & $0.3 \pm 0.1$ & $0.9 \pm 0.1$ & $3.9 \pm 0.6$ & $0.7 \pm 0.1$ & $0.3 \pm 0.1$ & \dots & $0.68 \pm 0.07$ & $0.21 \pm 0.05$ & $0.47 \pm 0.07$ & $0.45 \pm 0.04$\\
    Region 4  & $0.14 \pm 0.04$ & $0.56 \pm 0.03$ & $2.1 \pm 0.2$ & $0.26 \pm 0.04$ & $0.41 \pm 0.03$ & $0.029 \pm 0.008$ & $0.70 \pm 0.02$ & $0.12 \pm 0.01$ & $0.31 \pm 0.02$ & $0.44 \pm 0.02$\\
    Region 5  & \dots & $0.24 \pm 0.04$ & $0.79 \pm 0.07$ & $0.12 \pm 0.04$ & $0.21 \pm 0.03$ & $0.032 \pm 0.008$ & $0.45 \pm 0.02$ & $0.09 \pm 0.01$ & $0.22 \pm 0.01$ & $0.33 \pm 0.04$\\
    Region 6  & \dots & $0.26 \pm 0.03$ & $0.8 \pm 0.2$ & $0.10 \pm 0.03$ & $0.13 \pm 0.03$ & $0.029 \pm 0.008$ & $0.38 \pm 0.02$ & $0.07 \pm 0.01$ & $0.16 \pm 0.01$ & $0.29 \pm 0.02$\\
    Region 7  & $0.21 \pm 0.03$ & $0.62 \pm 0.03$ & $2.0 \pm 0.2$ & $0.30 \pm 0.04$ & $0.45 \pm 0.03$ & $0.058 \pm 0.008$ & $0.77 \pm 0.02$ & $0.18 \pm 0.01$ & $0.37 \pm 0.03$ & $0.47 \pm 0.04$\\
    Region 8  & \dots & $0.09 \pm 0.04$ & $0.22 \pm 0.07$ & $0.09 \pm 0.04$ & $0.10 \pm 0.03$ & $0.051 \pm 0.009$ & $0.16 \pm 0.02$ & \dots & \dots & $0.15 \pm 0.02$\\
    Region 9  & \dots & $1.3 \pm 0.1$ & $4.4 \pm 0.6$ & $0.9 \pm 0.1$ & $0.5 \pm 0.1$ & $0.09 \pm 0.03$ & $1.32 \pm 0.07$ & $0.51 \pm 0.05$ & $0.83 \pm 0.07$ & $0.6 \pm 0.1$\\
 \hline
 \label{PAHlinetable}
\end{tabular}\\
{$^a$Units are 10$^{-15}$~W~m$^{-2}$. Uncertainties are as calculated by PAHFIT (see text for details). \\
$^b$PAH complexes, as described in text.}
}
\end{minipage}
\end{table*}

\begin{table*}
 \centering
 \begin{minipage}{200mm}
 
\caption{PAH Emission Line Equivalent Widths$^a$}
 {\scriptsize
  \begin{tabular}{l c c  c  c  c  c  c  c  c  c c }
  \hline 
     {Region }&{5.7 $\mu$m  }&{6.2 $\mu$m  }&{7.7 $\mu$m$^b$  }&{8.3 $\mu$m  }&{8.6 $\mu$m  }&{10.7 $\mu$m  }&{11.2 $\mu$m$^b$  }&{12.0 $\mu$m  }&{12.7 $\mu$m$^b$  }&{17.0 $\mu$m$^b$  } \\
 \hline
       Bulge  & \dots & $1.2 \pm 0.2$ & $4.1 \pm 0.4$ & $0.51 \pm 0.07$ & $0.30 \pm 0.06$ & $0.03 \pm 0.01$ & $0.78 \pm 0.04$ & $0.22 \pm 0.02$ & $0.49 \pm 0.05$ & $1.16 \pm 0.04$\\
    Region 1  & $0.39 \pm 0.08$ & $1.2 \pm 0.1$ & $3.4 \pm 0.3$ & $0.43 \pm 0.04$ & $0.47 \pm 0.04$ & $0.09 \pm 0.01$ & $1.45 \pm 0.04$ & $0.43 \pm 0.03$ & $0.76 \pm 0.04$ & $1.26 \pm 0.05$\\
    Region 2  & $0.28 \pm 0.04$ & $1.02 \pm 0.06$ & $3.4 \pm 0.2$ & $0.32 \pm 0.04$ & $0.70 \pm 0.04$ & $0.07 \pm 0.01$ & $1.58 \pm 0.04$ & $0.35 \pm 0.02$ & $0.85 \pm 0.05$ & $1.33 \pm 0.06$\\
    Region 3$^c$  & $4.3 \pm 2.3$ & $8.3 \pm 2.3$ & $18.5 \pm 4.2$ & $2.8 \pm 0.8$ & $0.9 \pm 0.5$ & \dots & $2.1 \pm 0.3$ & $0.7 \pm 0.2$ & $1.6 \pm 0.3$ & $2.4 \pm 0.3$\\
    Region 4  & $0.28 \pm 0.09$ & $1.0 \pm 0.1$ & $3.7 \pm 0.4$ & $0.48 \pm 0.07$ & $0.77 \pm 0.07$ & $0.07 \pm 0.02$ & $1.67 \pm 0.06$ & $0.31 \pm 0.04$ & $0.82 \pm 0.06$ & $1.65 \pm 0.08$\\
    Region 5  & \dots & $0.12 \pm 0.03$ & $0.61 \pm 0.07$ & $0.10 \pm 0.03$ & $0.20 \pm 0.03$ & $0.05 \pm 0.01$ & $0.77 \pm 0.04$ & $0.17 \pm 0.03$ & $0.50 \pm 0.04$ & $1.6 \pm 0.2$\\
    Region 6  & \dots & $0.10 \pm 0.04$ & $0.6 \pm 0.2$ & $0.10 \pm 0.04$ & $0.14 \pm 0.03$ & $0.05 \pm 0.01$ & $0.77 \pm 0.05$ & $0.15 \pm 0.03$ & $0.42 \pm 0.05$ & $1.7 \pm 0.1$\\
    Region 7  & $0.32 \pm 0.05$ & $0.86 \pm 0.06$ & $2.2 \pm 0.1$ & $0.44 \pm 0.06$ & $0.69 \pm 0.06$ & $0.12 \pm 0.02$ & $1.81 \pm 0.07$ & $0.48 \pm 0.05$ & $1.08 \pm 0.09$ & $2.2 \pm 0.3$\\
    Region 8  & \dots & \dots & $0.10 \pm 0.05$ & $0.09 \pm 0.04$ & $0.10 \pm 0.03$ & $0.09 \pm 0.02$ & $0.30 \pm 0.04$ & \dots & \dots & $0.62 \pm 0.08$\\
    Region 9$^c$  & \dots & $\left(2.4 \pm 1.1\right) \times 10^{2}$ & $\left(1.4 \pm 0.4\right) \times 10^{2}$ & $15.5 \pm 5.6$ & $7.9 \pm 2.8$ & $0.5 \pm 0.2$ & $6.8 \pm 1.1$ & $2.3 \pm 0.6$ & $3.5 \pm 0.8$ & $2.2 \pm 0.8$\\
 \hline
 \label{EQW}
\end{tabular}\\
{$^a$Units are $\mu$m. Uncertainties are calculated using a Monte-Carlo method (see text for details). \\
$^b$PAH complexes, as described in text.\\
$^c$Continuum for Regions 3 and 9 is very weak.  Equivalent widths are highly uncertain and not considered in the analysis (see Section~\ref{sect:data_analysis}).}
}
\end{minipage}
\end{table*}

\subsection{PAH features}
\label{sect:pah}

PAHFIT returns fluxes and equivalent widths (EQWs) of PAH features, which are given in Tables~\ref{PAHlinetable} and \ref{EQW}. 
For easier comparison with previous work, the tables give measurements for the 7.7, 11.2, 12.7 and 17.0~$\mu$m PAH complexes
as defined by \citet{Smith:2007lr}, as computed from the individual constituent features measured by PAHFIT.
The intensities of the features do not include any contribution from the continuum, but the equivalent width computed by
\begin{equation}
{\rm EQW}=\int \frac{I_{\nu,{\rm feature}}}{I_{\nu, {\rm cont}}} \,d\lambda,
\end{equation}
is a measure of the ratio of the continuum emission ($I_{\nu, {\rm cont}} $) to the line strength 
($I_{\nu,{\rm feature}}=I_{\nu} - I_{\nu, {\rm cont}})$. 
The continuum emission is mainly from dust grains much larger than PAH molecules. Hence, by studying EQWs of PAHs, 
we can study how the PAHs compete with the dust grains in the mid-infrared wavelengths.  Uncertainties for line strengths are as calculated by PAHFIT's propagation of statistical uncertainties in the background-subtracted datacubes. Systematics and degeneracies in the fitting procedure are not included, and may dominate at higher signal-to-noise \citep{Smith:2007lr}.
Uncertainties for the EQWs for each PAH 
feature are calculated using a Monte-Carlo method. In that method, for each region, PAHFIT was run 500 times on 
randomly generated data points  normally distributed within the uncertainties of the spectrum. PAHFIT returned 500 EQWs for each 
PAH feature, and the standard deviation of EQWs for a given feature was taken as its uncertainty. 
The EQWs from Regions 3 and 9 were removed from further analysis because the negligible dust continuum for these spectra makes
the EQWs highly uncertain.

\subsection{Atomic line features}
\label{sect:atomic}

\begin{figure}
\centering
\includegraphics[scale=0.37]{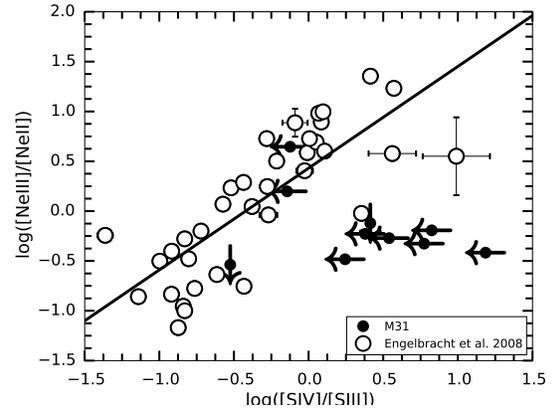}
\caption{ Log([Ne~{\sc iii}]/[Ne~{\sc ii}])  vs log([S~{\sc iv}]/[S~{\sc iii}])  for the M31 regions in our sample (black dots) and  the starburst sample from \citet{Engelbracht_2008} (open dots). The straight line is the line of best fit for the starburst sample.
Arrows show $3\sigma$ upper limits.
}
\label{SvsNe}
\end{figure}

\begin{table*}
 \centering
 \begin{minipage}{100mm}
 
\caption{Atomic Emission Line Strengths$^a$}
  \begin{tabular}{l c c  c  c  c  c  }
  \hline
  {Region  }&{[Ar~{\sc ii}] }&{[Ar~{\sc iii}]  }&{[S~{\sc iv}]}&{[Ne~{\sc ii}]   }&{[Ne~{\sc iii}]   }&{[S~{\sc iii}]  }\\
{}&{\tiny{7.0 $\mu$m} }&{\tiny{9.0 $\mu$m }}&{\tiny{10.5 $\mu$m}}&{\tiny{12.8 $\mu$m  }}&{\tiny{15.5 $\mu$m } }&{\tiny{18.7 $\mu$m }} \\
 \hline 
       Bulge  & $<0.12$ & $0.17 \pm 0.03$ & $<0.11$ & $0.07 \pm 0.01$ & $0.025 \pm 0.006$ & $0.007 \pm 0.003$\\
    Region 1  & $<0.05$ & $<0.06$ & $0.020 \pm 0.004$ & $0.020 \pm 0.004$ & $<0.01$ & $0.008 \pm 0.001$\\
    Region 2  & $<0.05$ & $<0.06$ & $<0.02$ & $0.022 \pm 0.004$ & $<0.01$ & $0.003 \pm 0.002$\\
    Region 3  & $<0.15$ & $0.09 \pm 0.02$ & $<0.10$ & $0.03 \pm 0.01$ & $0.020 \pm 0.004$ & $0.015 \pm 0.003$\\
    Region 4  & $<0.04$ & $<0.04$ & $<0.02$ & $<0.02$ & $<0.01$ & $0.004 \pm 0.002$\\
    Region 5  & $<0.04$ & $<0.02$ & $<0.02$ & $<0.01$ & $0.009 \pm 0.004$ & $0.008 \pm 0.004$\\
    Region 6  & $0.02 \pm 0.01$ & $0.014 \pm 0.007$ & $<0.01$ & $<0.01$ & $0.019 \pm 0.002$ & $0.019 \pm 0.002$\\
    Region 7  & $<0.04$ & $<0.04$ & $0.008 \pm 0.003$ & $0.035 \pm 0.005$ & $<0.01$ & $0.027 \pm 0.004$\\
    Region 8  & $<0.04$ & $0.017 \pm 0.006$ & $<0.02$ & $<0.01$ & $0.041 \pm 0.002$ & $0.023 \pm 0.002$\\
    Region 9  & $0.09 \pm 0.04$ & $0.12 \pm 0.03$ & $<0.09$ & $0.13 \pm 0.01$ & $<0.04$ & $0.05 \pm 0.02$\\
\hline
 \label{Atomic}
\end{tabular}\\
{ $^a$Units are 10$^{-15}$~W~m$^{-2}$. Uncertainties are as calculated by PAHFIT (see text for details). Upper limits ($3\sigma$) are indicated with a $<$ mark.  }
\end{minipage}
\end{table*}

PAHFIT also returns the line strengths and uncertainties for atomic lines, listed in Table~\ref{Atomic}.
Not all lines were detected by PAHFIT, so we calculated upper limits for non-detected lines.%
\footnote{To find  the upper limits for the flux of missing atomic lines, we assumed the line to be a 
Gaussian profile with a FWHM as given by PAHFIT. The peak intensity was taken to be 3 times the RMS, where RMS is the root mean square of 
the noise at the position of a missing line.}
Line ratios of [Ne~{\sc iii}]/[Ne~{\sc ii}] and [S~{\sc iv}]/[S~{\sc iii}]~18 have been used as an indication of the radiation hardness, and
\citet{Engelbracht_2008} defined a combination of these two line ratios as a `radiation hardness index (RHI)':
\begin{equation}
{\rm RHI} = \left( \log\frac{\textrm{[Ne~{\sc iii}] }}{\textrm{[Ne~{\sc ii}]}} + \left[0.71 + 1.58\log\frac{\textrm{{[S~{\sc iv}]}}}{\textrm{{[S~{\sc iii}]}}}\right]\right) /2
\label{eq:rhi}
\end{equation}
Here, 1.58 and 0.71 are the slope and intercept of the [Ne~{\sc iii}]/[Ne~{\sc ii}]  vs [S~{\sc iv}]/[S~{\sc iii}] plot (Figure \ref{SvsNe}) for the starburst sample from 
\citet{Engelbracht_2008}. 
Figure \ref{SvsNe}  compares the atomic line emission from the selected regions of M31 to the starburst galaxy sample;
although none of our spectra has detections of all four lines, our limits are mostly consistent with the trend.
To compute RHI in the case of missing lines, we used the term in Equation~\ref{eq:rhi} 
for which both lines are detected. For regions with only one line detected per element, we computed RHI upper limits 
using Equation~\ref{eq:rhi}.

\section{Results: extranuclear regions}

\subsection{PAH band ratios}
\label{sect:pah_ratios}

\begin{figure}
\centering
\includegraphics[scale = 0.4]{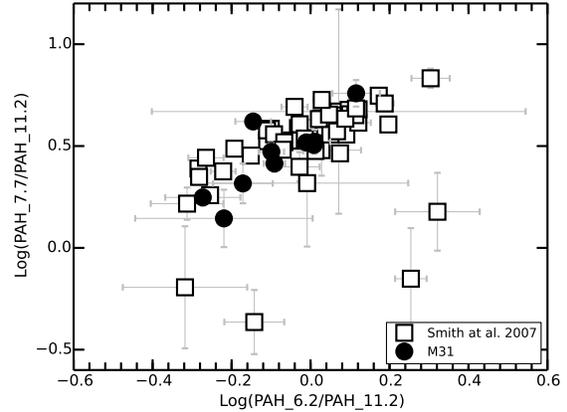}
\caption{Ratios of PAH feature fluxes (7.7~$\mu$m/11.2~$\mu$m versus 6.2~$\mu$m/11.2~$\mu$m) for 10 regions in M31.
Open squares represent the central regions of nearby galaxies as observed in the SINGS sample by \citet{Smith:2007lr}.
}
\label{PAHlines}
\end{figure}

Both the 6.2 and 7.7~$\mu$m features are thought to be coming from ionized PAHs and the 11.2~$\mu$m feature from neutral PAHs. Therefore we expect to see a correlation between the intensities of 6.2 and 7.7~$\mu$m PAH features normalized by the 11.2~$\mu$m feature.  Figure \ref{PAHlines}  compares the PAH flux ratios of 7.7/11.2  and 6.2/11.2 features. The figure shows a good correlation between these two PAH line ratios, consistent with that of the SINGS sample shown by \citet{Smith:2007lr}.
A similar correlation was also reported by  \citet{Galliano2008} for a sample of galaxies and a handful of extended H{\sc ii} regions
and by \citet{Vermeij2002} for Galactic and Magellanic Cloud H{\sc ii}\,regions. This provides evidence that the PAH emission from M31 is not unusual.

\begin{figure*}
\centering
\includegraphics[scale=0.85]{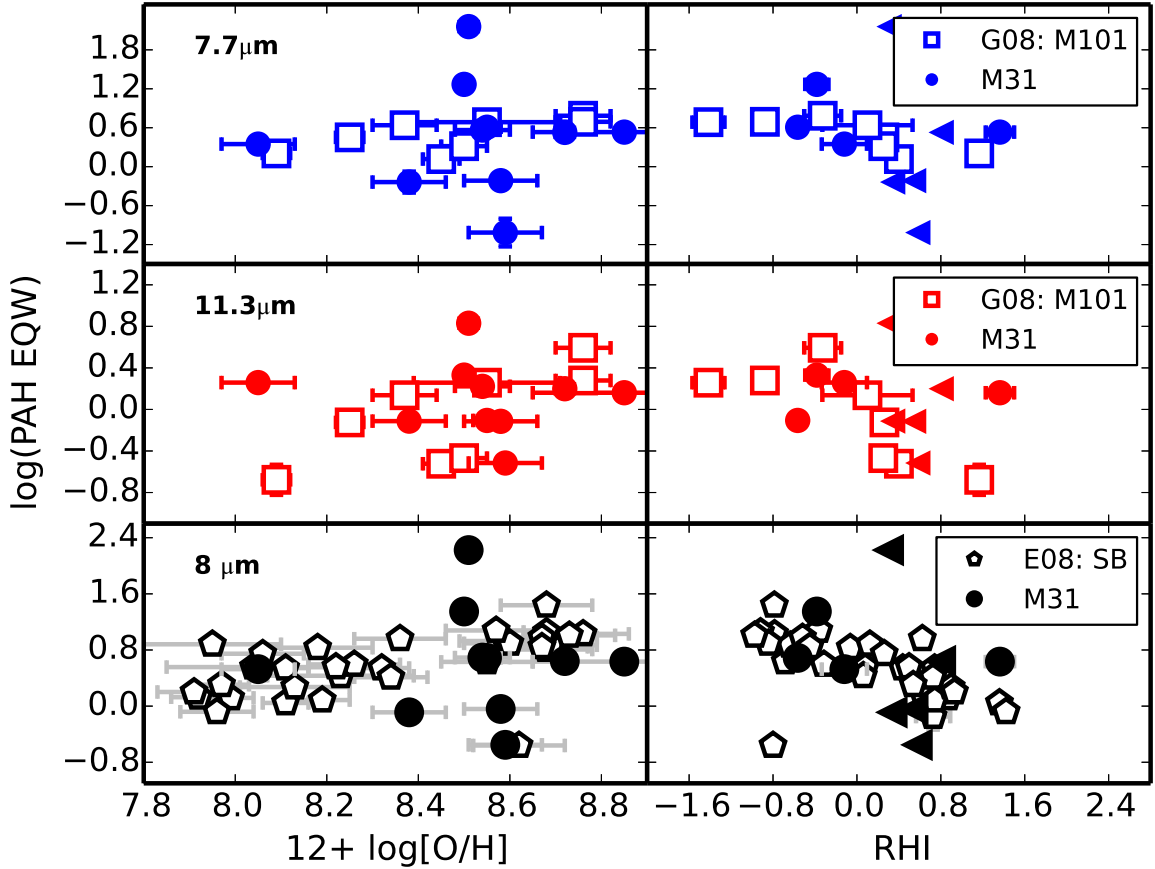}
\caption{Equivalent width of  PAH feature versus metallicity (left panels) and radiation hardness index (RHI; right panels).
Equivalent  widths are not normalized; metallicities of the M31 regions have had 0.35~dex subtracted to account for the offset  between direct and strong-line measurements \citep{Croley2015}. 
Solid points: M31 sample; triangles represent upper limits. Open points:  starburst galaxy sample from \citet[bottom panels]{Engelbracht_2008} or
 H~{\sc ii} regions in M101 observed  by \citet[middle and top panels]{Gordon:2008lr}.  
}
\label{rhi_met_eqw}
\end{figure*}

\subsection{PAH equivalent widths versus radiation hardness and metallicity}
\label{sect:eqw_rh}

As mentioned in the introduction, PAH equivalent widths tend to decrease as radiation hardness increases,
and as metallicity decreases \citep{Calzetti:2010fk}.  

Figure~\ref{rhi_met_eqw} shows the equivalent widths of the  8~$\mu$m feature 
(a combination of the 7.7, 8.3 and 8.6~$\mu$m PAHFIT components) and the 
 7.7 and 11.2~$\mu$m features as a function of both metallicity and RHI. The starburst galaxy sample of \citet{Engelbracht_2008} 
and  H{\sc ii} regions in M101 \citep{Gordon:2008lr} are plotted for comparison.
The M31 regions cover approximately the same range of measurements as the starburst galaxies and M101 H{\sc ii} regions;
the outliers are regions 3 and 9 with high equivalent widths (uncertain due to low continuum, see Section~\ref{sect:pahfit}) and
region 8 with low equivalent widths (noisy spectrum as well as substantial modelled contribution from starlight, see Figure~\ref{PAHFITplots}).
No clear trend is defined by the M31 data, unsurprising given the uncertainties and the limited
number  of regions.
We do not have enough data from low-metallicity regions in M31 to observe the expected decrease of PAH EQW with decreasing 
metallicity; however the M31 data occupy the same region of parameter space as the M101 data
and we conclude that the EQWs of the regions in M31 are consistent with previously published ``normal'' values.

\section{Results: the nuclear region}
\label{sect:nucleus}

\begin{figure*}
\centering
\resizebox{\hsize}{!}{%
\includegraphics{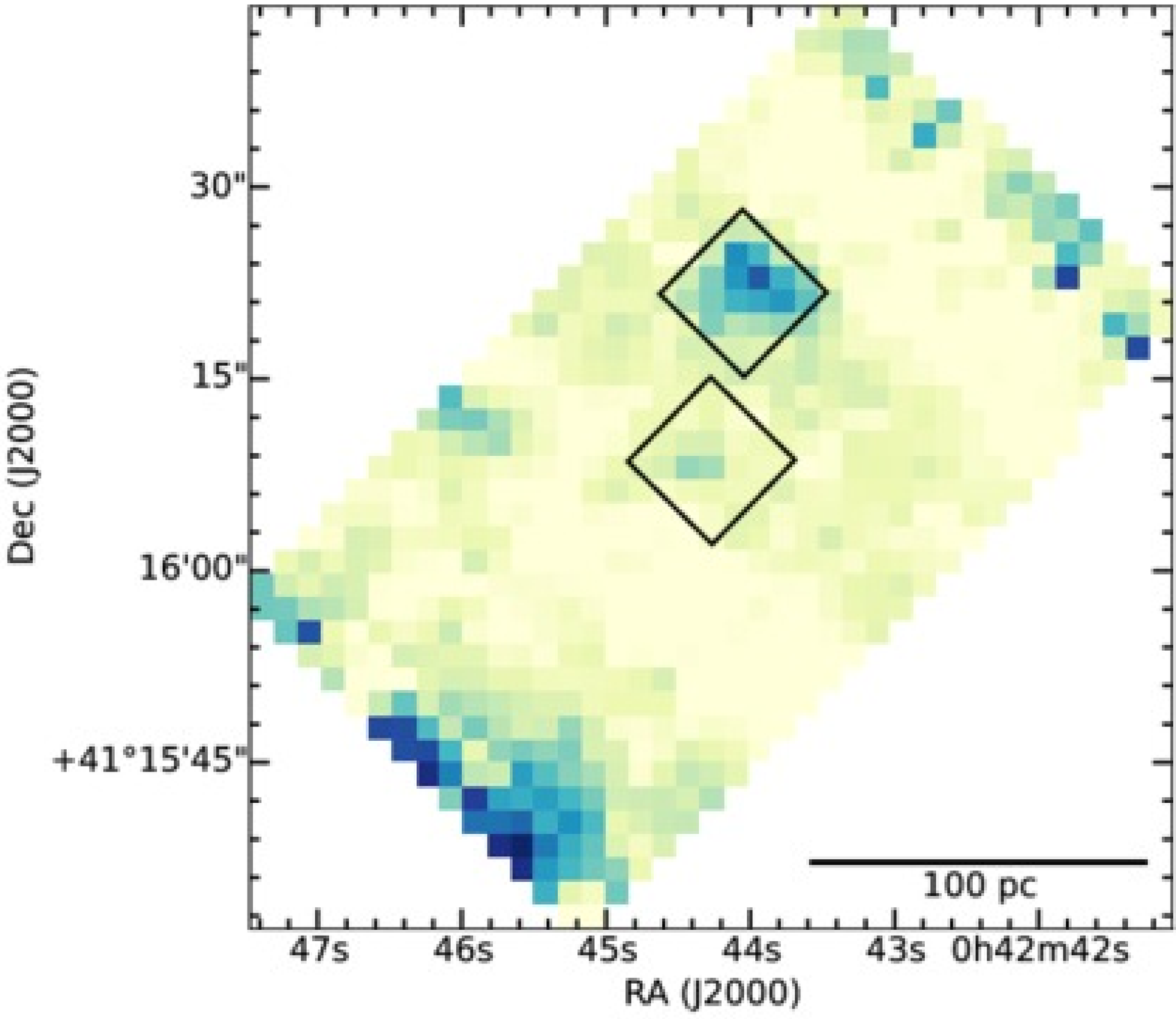}
\includegraphics{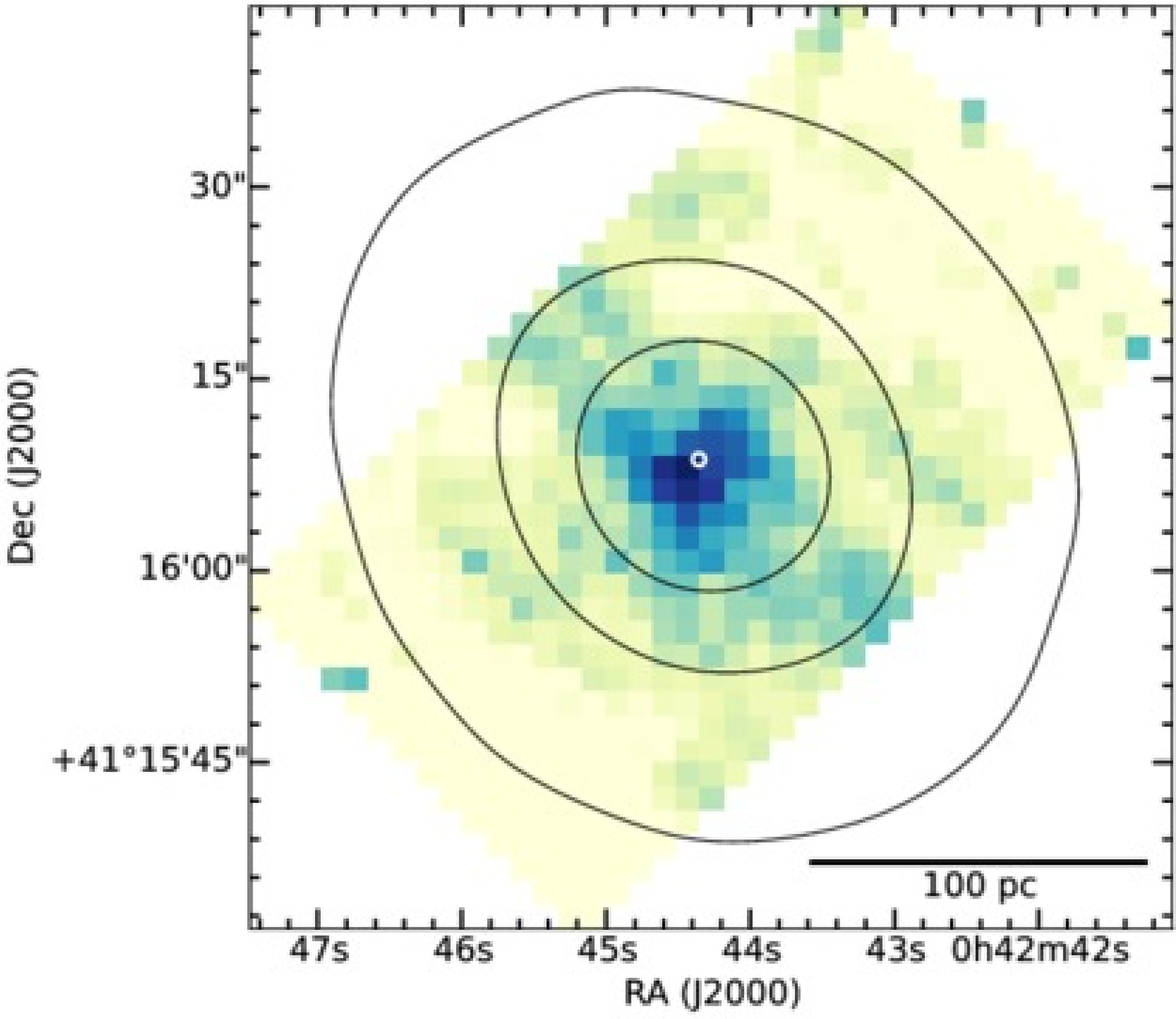}}
\caption{
Integrated strength of the 11.2~$\mu$m PAH emission (left) and the silicate emission (from 9 to 11~$\mu$m, continuum subtracted; right) 
around the nucleus of M31. 
The black boxes in the left panel are the 9\arcsec\ $\times$ 9\arcsec\ sub-apertures (centre and north region) used to extract spectra.  
The small white circle in the right panel shows a position halfway between 
the two components of the nucleus \citep[coordinates for P1 and P2/P3 in Table 1 of][]{NucleusREF}.
These components are separated by 0\farcs5, which is well below the IRS spatial resolution.
The contours in the right panel represent
the distribution of 3.6~$\mu$m luminosity, indicating the orientation of the M31 bulge.
}
\label{nuc11}
\end{figure*}

\begin{figure*}
\centering
\includegraphics[height = 10 cm]{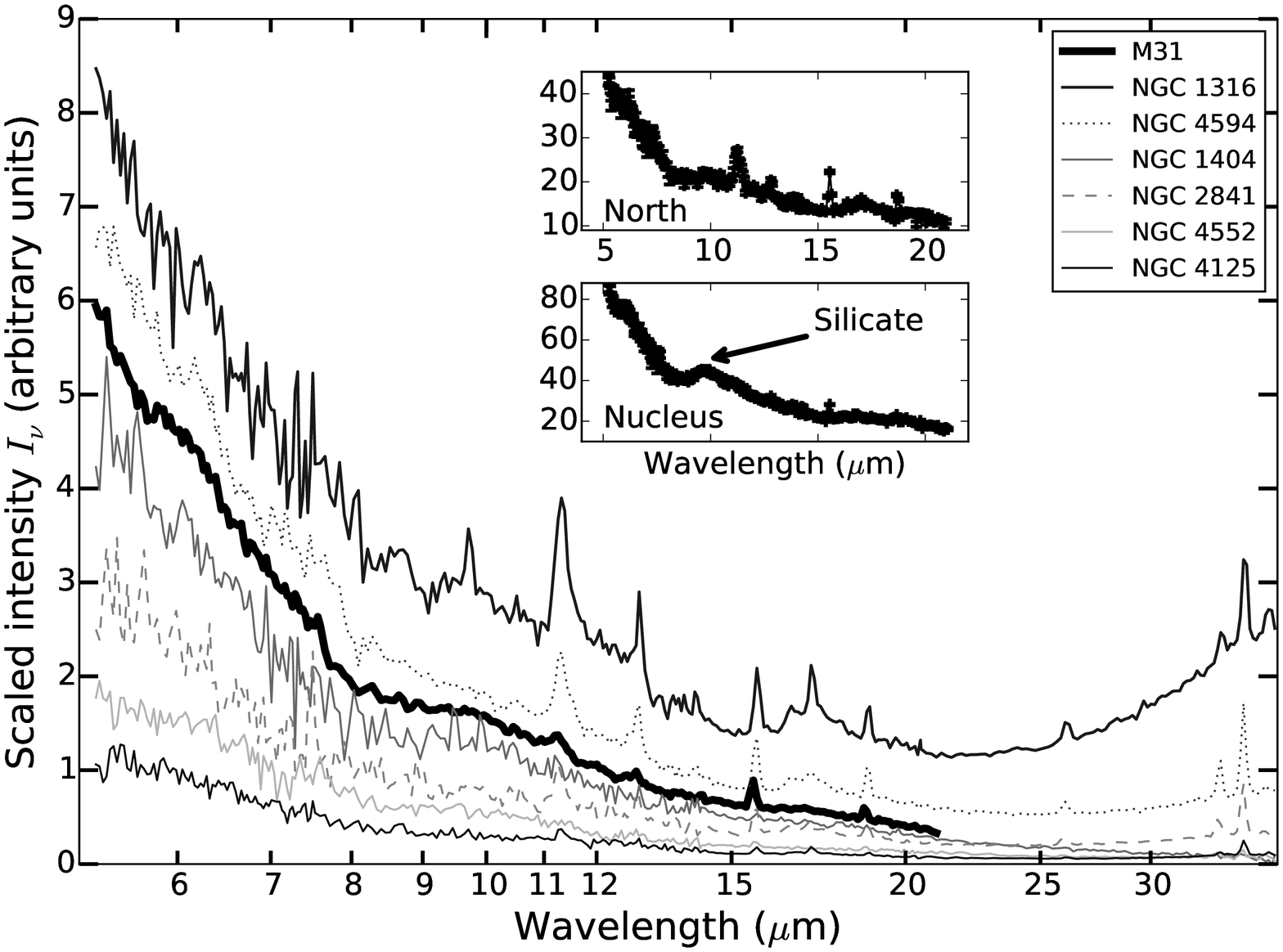}
\caption{Mid-infrared spectrum of the nucleus of M31 (thick line; full spectrum) over-plotted with spectra extracted close to the nuclei of 6 nearby galaxies 
which have similar spectral shapes \citep{Smith:2007lr}. The comparison spectra are offset by arbitrary amounts for visibility; they follow the same vertical order
as in the legend.
NGC 4552, NGC 1404 and NGC 4125 are elliptical galaxies and NGC 4594 and NGC 2841 are spiral galaxies. 
NGC 1316 is a lenticular galaxy. The inset shows the spectra extracted from the centre region of the M31 nucleus  and from the north region 
shown in Figure \ref{nuc11}.}
\label{smithspec}
\end{figure*}

Examining the {\em Spitzer}-IRS spectral data cube for the nuclear region, we noticed that different spectral features vary spatially within this region.
The 11.2~$\mu$m PAH emission is discrete and patchy (Figure \ref{nuc11}, top).  
Indeed, the majority of the 11.2~$\mu$m  PAH emission is from a region centered at (00:42:43.947, +41:16:22.92), 15\arcsec\ north of the nucleus and not from 
the nucleus itself. Weaker 11.2~$\mu$m PAH emission is also found near the edge of the map peaking at 00:42:45.497, +41:15:43.97 and near 00:42:45.869; +41:16:11.38.
The locations of the two weaker 11.2~$\mu$m PAH emission peaks are also near positions exhibiting CO(2-1) line emission \citep[\#36 and 28 of][the strongest 11.2~$\mu$m PAH emission peak is outside the CO FOV]{Melchior2013}. 
On the other hand, the centre shows no PAH emission, but it does have clear silicate emission around 9.7~$\mu$m (Figure \ref{nuc11}, bottom).%
\footnote{The spatial resolution and pixel scale of the ISOCAM data are not sufficient to resolve the centre  and north regions.}
Although [NeIII] 15.5~$\mu$m line emission is strong at the three locations with 11.2~$\mu$m PAH emission and weak(er) at the nucleus, 
it is also present across the nuclear region and thus spatially distinct from the silicate and PAH emission. 

Some structures seen in the optical near the M31 nucleus, such as the double nucleus and cluster of young stars, cannot be resolved with the data discussed here. The IRS spectral cubes have pixel sizes of 1\farcs8, and the SL PSF FWHM is 2.5--3\arcsec ~depending on the wavelength, while the two components of the nucleus are separated by only 0\farcs5 \citep{Bender2005}. However, the north 11.2~$\mu$m  PAH emission appear to be marginally spatially resolved: its radial profile has a full width at half maximum (FWHM) of 5--7\arcsec\ (corresponding to 19--27~pc).

\begin{figure}
\centering
\includegraphics[width = 8.6 cm]{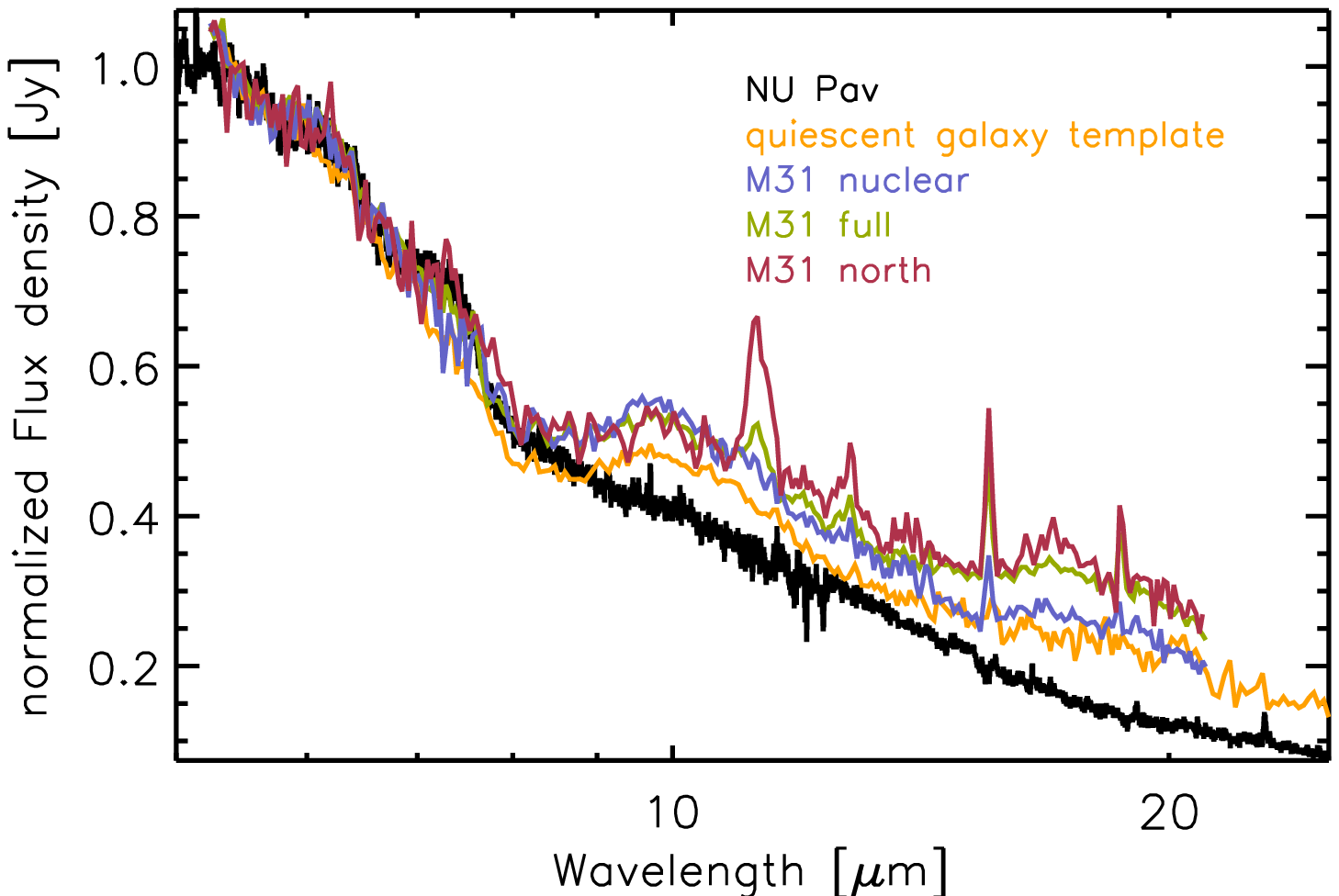}
\caption{The spectra of the M31 nucleus (nuclear, north and full) compared with the M giant NU Pav and the quiescent galaxy template of \citet{Kaneda:08}, normalized to the 5.4 $\mu$m flux density. The legend gives the objects from lowest to highest flux at the longest wavelengths.}
\label{fig:nuc_stellar}
\end{figure}

To examine this spatial variation towards the nucleus in more detail, we extracted spectra from the centre and the North regions using  $9\arcsec \times 9\arcsec$ square apertures in addition to the full 30\arcsec $\times$ 50\arcsec\ M31 nuclear spectrum (further referred to as nuclear spectrum, north spectrum and full spectrum respectively), as shown in Figures~\ref{smithspec} and~\ref{fig:nuc_stellar}. Both spectra show a blue continuum and atomic fine-structure lines of Ne and S, but they exhibit distinct dust emission consistent with the spatial maps: PAH emission is detected in the north spectrum while silicate emission is dominating in the nuclear spectrum.  In addition, weak H$_2$ emission is present at 17~$\mu$m in the north spectrum.\\

To investigate the stellar contribution to these spectra, we made a comparison with the IR spectra of NU Pav, an M giant \citep{Sloan:15} and the template spectrum of quiescent elliptical galaxies as constructed by \citet{Kaneda:08} to investigate the PAH emission in elliptical galaxies (in particular the 7.7 $\mu$m PAH band). The IR spectrum of NU Pav is very similar to this template spectrum, but it lacks silicate emission and dust continuum emission at the longer wavelengths. 
Assuming that the NIR fluxes are completely due to the stellar component, we can estimate the contribution of the stellar component to the 5.4 $\mu$m flux by normalizing the spectrum of NU Pav and the full M31 spectrum to the H band flux \citep[following][]{Vega:10}. NIR fluxes for NU Pav are taken from \citet{Gezari:99} and for M31 are determined by integrated photometry of the 2MASS Large Galaxy Atlas \citep{Jarrett:03} images over the spectral extraction regions. This allows us to determine the stellar contribution to the 5.4 $\mu$m flux of the full M31 spectrum, which is found to be 96\%. Therefore, in order to compare with the quiescent galaxy template, we further assume that 100\% of the flux at 5.4 $\mu$m is stellar; Figure~\ref{fig:nuc_stellar} shows a comparison of the spectra when normalized at 5.4 $\mu$m. The match between the spectra up to $\sim$ 8 $\mu$m is very close, with no evidence for non-stellar emission in the M31 spectra. Furthermore, it is striking that the nuclear, full and north spectra of M31 match each other extremely well up to $\sim$14 $\mu$m, including the 9 $\mu$m silicate emission,  except for the PAH emission and fine-structure line emission. In contrast, the 9 $\mu$m silicate emission is less intense in the quiescent galaxy template and absent in NU Pav; both of these sources also lack PAH emission and fine-structure line emission. Given that the presence and strength of the silicate emission, if stellar, depends on the age of the stellar population \citep{Villaume:15}, this resemblance suggests that the silicate emission is also stellar in nature. These authors predict the presence of silicate emission for stellar population ages of $< 10^{8.5}$ yr.  M31 is known to have stars of this age group: a 200 Myr-old starburst in the central 1$\arcsec$ plus an intermediate-age population both at the centre and the inner bulge \citep{Bender2005, Saglia:10, Dong:15}.  Hence, the nuclear spectrum is largely dominated by the stellar component with little contribution from the nuclear region itself, which is mostly manifested by the fine-structure line emission. Indeed, while a weak dust continuum flux may be present at the longer wavelengths, the flux increase around $\sim$ 15 $\mu$m is likely due to the presence of the 18 $\mu$m silicate emission band. Therefore, assuming that the spectral shape of the stellar contribution does not change spatially within the M31 nuclear region, we can use the nuclear spectrum as a template of the stellar contribution to the north and full spectra. \\

\begin{figure*}
\centering
\includegraphics[width = 8 cm]{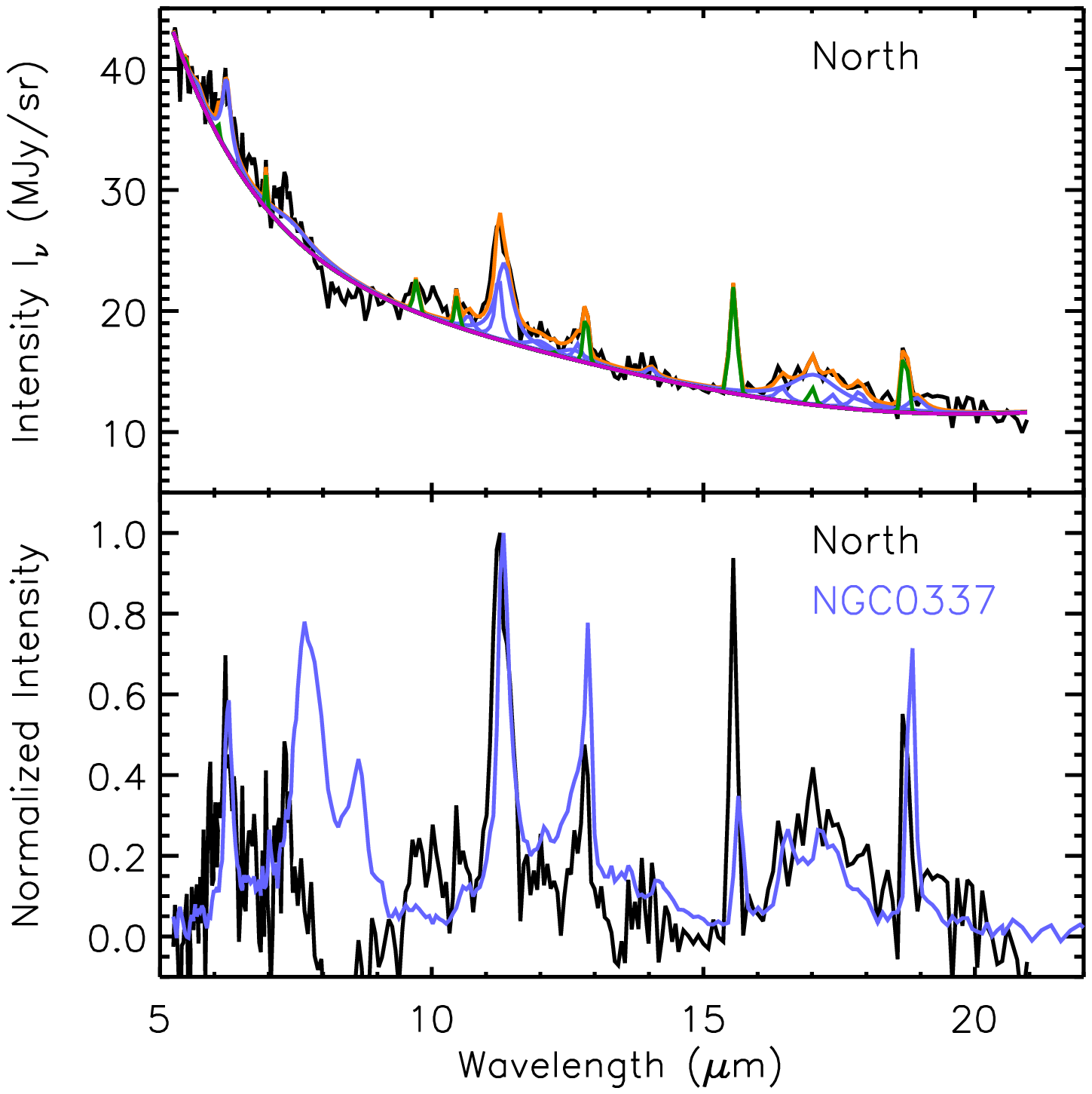}
\includegraphics[width = 8 cm]{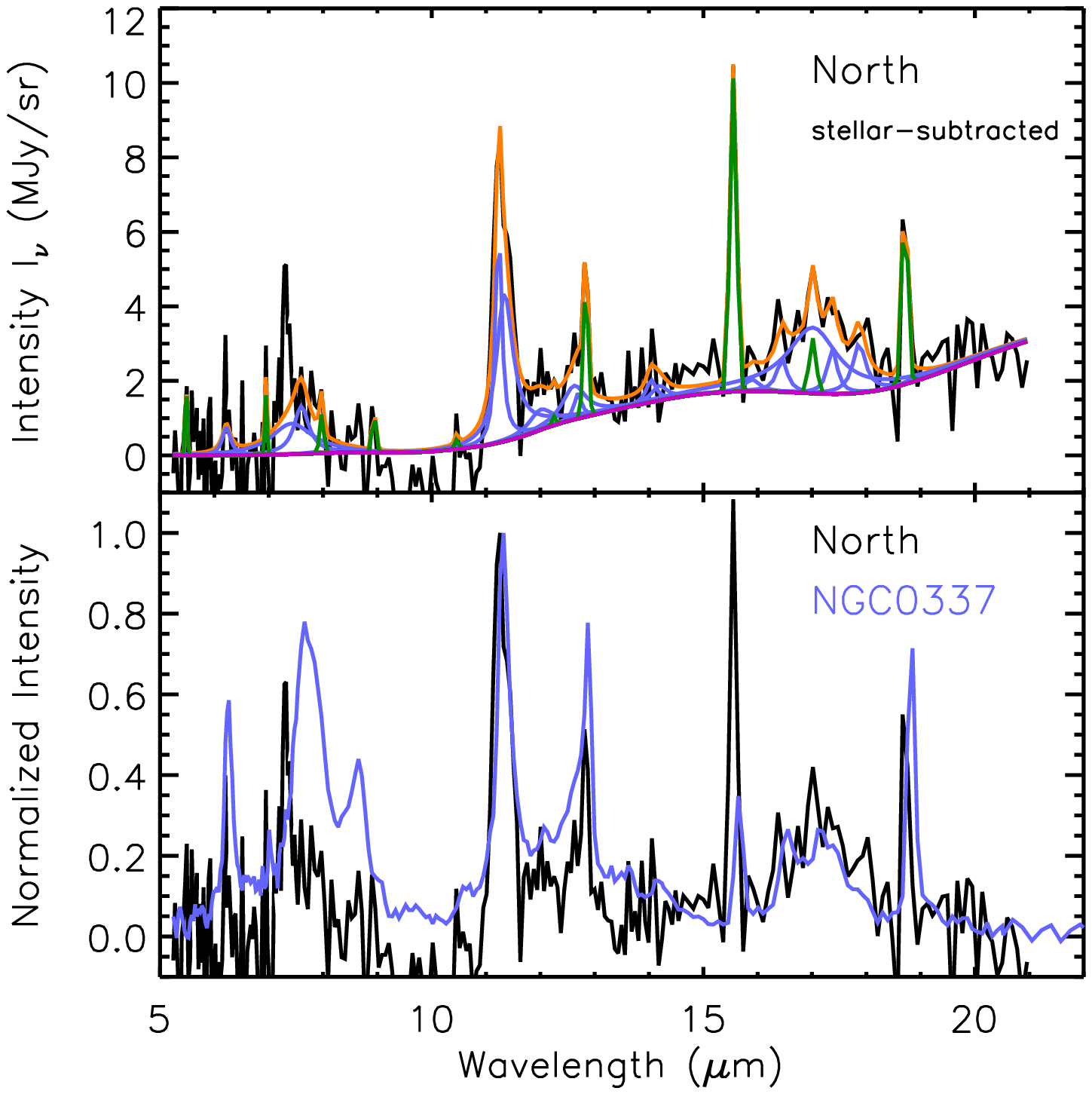}
\caption{(top) PAHFIT result for the M31 north spectrum (left) and the M31 north spectrum subtracted by the stellar component (as defined by the M31 nuclear spectrum, right): fit (orange), continuum (magenta), individual dust components (blue), individual fine-structure lines and H$_2$ lines (green). (bottom) Continuum subtracted spectra of the north region (left) and the north region without the stellar component (right) compared with that of the HII-type galaxy NGC0337 \citep{Smith:2007lr}, normalized to the peak intensity of the  11.2~$\mu$m PAH emission. }
\label{fig:nuc_pahfit}
\end{figure*}

The north spectrum exhibits clear PAH emission, in particular at 11.2 and 15--20 $\mu$m (Figure~\ref{fig:nuc_stellar}). However, the typical 6--9 $\mu$m PAH emission is very weak or even absent. Such unusual PAH emission has also been reported towards elliptical galaxies \citep[e.g.][]{Kaneda:05, Bregman:08, Kaneda:08, Vega:10} and in some low-luminosity active galactic nuclei \citep[LLAGN, ][]{Smith:2007lr}. \citet{Bregman:08} reported however that the shape of the stellar emission greatly influences the appearance, and thus apparent strength, of the 6-9 $\mu$m PAH emission. Therefore, we apply two different corrections for the stellar component to the north spectrum in order to investigate the `true' PAH emission. First, we apply PAHFIT \citep{Smith:2007lr} to the spectrum of the north region and rely on the stellar component within PAHFIT (a blackbody of 5000K) to account for the stellar contribution to the spectrum (Figure~\ref{fig:nuc_pahfit}, left). This stellar component clearly does not take into account possible stellar spectral features such as silicate emission, H$_2$O absorption (at 6.6 $\mu$m), SiO absorption (at 8 $\mu$m) etc.  Secondly, we first subtract the stellar contribution by using the nuclear spectrum (normalized at 5.4 $\mu$m) and then apply PAHFIT to this stellar-subtracted north spectrum (Figure~\ref{fig:nuc_pahfit}, right).

For both methods, the continuum emission is overestimated between $\sim$ 8 to 11 $\mu$m indicating that the silicate emission may be slightly overestimated when using the nuclear spectrum and confirming that a 5000~K blackbody (the stellar component in PAHFIT) is only a first-order approximation for the stellar contribution. In both cases, we further subtract the total continuum emission as found by PAHFIT and compare the residuals to a typical PAH emission spectrum as exemplified by the HII-type galaxy NGC 0337 \citep{Smith:2007lr}. Relative to the 11.2 $\mu$m PAH emission, the 15-20 $\mu$m emission is similar for both the north region and a typical PAH emission spectrum while the 6--9 $\mu$m PAH emission remains distinct in the north region. However, the 6--9 $\mu$m PAH emission differs for the two methods applied. When using the stellar component of PAHFIT, a 6.2 $\mu$m PAH emission is present while no significant 7.7 and 8.6 $\mu$m emission is present. In contrast, when using the normalized nuclear spectrum as the stellar component, the 6.2 and 8.6 $\mu$m PAH emission is largely absent, and the 7.7 $\mu$m PAH emission is weak. This is confirmed by the PAH intensity ratios found by PAHFIT: the 7.7/11.2 $\mu$m PAH intensity ratio is 0.79 for the M31 north spectrum which is much lower than the average in the SINGS sample \citep[3.6,][]{Smith:2007lr}. As mentioned, the two methods do not agree regarding the 6.2 $\mu$m band, resulting in 6.2/11.2 $\mu$m PAH intensity ratios of 1.18 and 0.14 (for the PAHFIT and nuclear stellar component respectively) whereas the SINGS average is 1.1 \citep{Smith:2007lr}. In contrast, the 17/11.2 $\mu$m PAH intensity ratios are 0.46 and 0.52 (for the PAHFIT and nuclear stellar component respectively), which is consistent with the SINGS average of 0.53 \citep{Smith:2007lr}. 

The nuclear component has the largest influence on the 7--9 $\mu$m PAH emission due to the presence of silicate emission and SiO absorption while the 6.2 $\mu$m PAH emission is less affected (due to weaker H$_2$O absorption). Hence, it is remarkable that we recover a 'normal' 6.2 $\mu$m PAH emission by using a stellar blackbody while it is largely absent when a more `appropriate' nuclear correction is applied. This testifies to the sensitivity of the deduced `true' PAH emission on the applied correction for the stellar component. Although the 7.7 $\mu$m PAH emission is partly recovered after the stellar component correction, the 7.7/11.2 PAH intensity ratio is still well below the average for the SINGS sample for both applied methods. Hence, until more accurate stellar templates are available, it is fair to conclude that the north spectrum does not show typical PAH emission. This is in contrast with the results of \citet{Bregman:08} who recover a normal PAH spectrum for the elliptical galaxy NGC4697 after subtraction of a quiescent elliptical galaxy spectrum. But it is consistent with \citet{Kaneda:08} who found that 80\% of elliptical galaxies in their sample (of 15) remain anomalously weak 7.7 $\mu$m emission after correction for the stellar component.  \\

As mentioned earlier, similar atypical PAH emission has been detected towards elliptical galaxies \citep[e.g.][]{Kaneda:08, Vega:10} and in some low-luminosity active galactic nuclei (LLAGN) reported by \citet{Smith:2007lr}.  To place M31 in the context of these other galaxies,
Figure \ref{smithspec} compares the full 30\arcsec $\times$ 50\arcsec\ M31 nuclear spectrum with the nuclear spectra of similar spectral shape from six SINGS galaxies \citep{Smith:2007lr}.%
\footnote{The IRS spectra for the SINGS galaxies were extracted over areas ranging from 2 to 8 kpc$^2$, whereas the M31 spectrum covers 0.02~kpc$^2$.}
All 6 SINGS galaxies share similar PAH feature characteristics to the M31 spectrum (e.g. weak 6--9~$\mu$m features), but none of them contains obvious silicate emission. 
The SINGS galaxies with similar spectral shapes include  three elliptical galaxies, two spirals, and a lenticular; there is some disagreement over the exact nuclear spectral types of these six galaxies \citep{kennicutt03,Smith:2007lr, moustakas2010}.  All are classified as some form of LLAGN
such as Seyfert or LINER \citep[luminous AGNs were intentionally omitted from the SINGS sample;][]{kennicutt03}, although they are
by no means the only LLAGNs in the SINGS sample.
\citet{Li09} concluded that the central black hole in M31 (M31*) is currently inactive, with direct observational signatures seen only
at radio and X--ray wavelengths, so finding additional signatures in the mid-infrared is of great interest.
To our knowledge, no such signatures have been reported; broadband mid-infrared imaging of the central 
regions of M31 \citep{davidge06,Barmby2006lr} did not identify unambiguous nuclear emission. The bluest
part of the spectrum in Figure~\ref{smithspec} is dominated by the continuum, in agreement with the
expectation that stellar light dominates at these wavelengths.

Could radiation from M31* be responsible for the suppression of the  6--8~$\mu$m PAH features compared
to the 11.3~$\mu$m feature?
As discussed by  \citet{Smith:2007lr} and \citet{Smith2010}, inferring such a suppression must be done with caution, 
because the 6--8~$\mu$m features are more susceptible to dilution by the stellar continuum. 
Several connections between PAH suppression and the presence of an AGN are possible, including destruction of small PAH molecules by a hard radiation field, modification of the structure of the PAH molecules, or weak ultraviolet continuum from low star formation rates 
leading to decreased PAH excitation \citep{Smith:2007lr, Diamond2010}.  In the latter case, the AGN is not the cause of the suppressed  6--8~$\mu$m features but rather is only detectable because the nuclear star formation rate is low.
Previous work has found low rates of star formation in the centre of M31: although \citet{Melchior2013} found a significant 
amount of cold gas in the centre of the galaxy, this gas does not appear to be associated with current star formation \citep[see also][]{Li09}.
In modelling the far-infrared spectral energy distribution, \cite{Groves2012} found that  
the old stellar population in the M31 bulge is sufficient to heat the observed dust; no young stellar population is needed. As shown, the nuclear spectrum is dominated by the stellar component and the dust continuum may be only weakly  present at the longer wavelengths. This is consistent with a low star formation rate. We conclude that PAH feature ratios cannot provide direct evidence for radiation from M31*.

\section{Summary and conclusions}
\label{sect:summary}
{\em Spitzer}/IRS spectral maps of 12 regions within M31 cover wavelengths 5--21~$\mu$m. 
The spectra from those regions, except for the nucleus, are similar to spectra obtained from other nearby  star-forming galaxies. 
Early  ISOCAM observations towards 4 regions of M31 showing a suppression 
of the 6--8~$\mu$m features' strength and an enhancement of  the 11.3~$\mu$m feature intensity and FWHM \citep{1998Cesarsky} were likely affected by the incorrect background subtraction methods applied. Indeed, both the re-processed ISOCAM data and the IRS data (except for the nucleus) exhibit typical PAH emission. 

We recover the well-known strong correlation between the 6.2/11.2 and 7.7/11.2 $\mu$m PAH intensity ratios that is governed by the PAH charge balance.  This is another indicator that the PAH emission in all regions but the nucleus is typical. The PAH EQWs in M31 regions do not show a clear decreasing trend with increasing radiation hardness, but are consistent with previous 
results from other nearby galaxies. The distribution of PAH EQWs with metallicity is well within the range of the starburst galaxy sample of \citet{Engelbracht_2008}. 
We did not have enough data from low-metallicity regions of M31 to observe the decreasing trend of EQWs at low metallicities which is visible in other galaxies.

Different spectral features (11.2~$\mu$m PAH emission, [NeIII] 15.5~$\mu$m line emission, silicate emission) show distinct spatial distributions in the nuclear region. The mid-infrared spectrum from the nucleus of M31 shows a strong blue continuum and clear silicate emission. This nuclear spectrum is largely dominated by the stellar component with little contribution from the nucleus itself. In addition to a blue continuum and silicate emission, the mid-infrared spectrum of a region north of the nucleus  (15\arcsec ~off-nucleus) exhibits suppressed 6--8~$\mu$m PAH features and normal 11.3 and 17.0 $\mu$m PAH emission.
This atypical PAH emission is only seen elsewhere towards some elliptical galaxies and low luminosity AGNs \citep[e.g.][]{Smith:2007lr, Kaneda:08, Vega:10}.

\section*{Acknowledgements}
The authors would like to thank the referee S. Hony whose comments have helped to improve the paper.  We also thank D. Stock, K. Sandstrom, S. Lianou for fruitful discussions and technical support
and S. Gallagher and G. Sloan for helpful comments. We thank H. Kaneda for providing a template spectrum of a quiescent elliptical galaxy. 
We acknowledge support from NSERC Discovery Grants to PB and EP, an NSERC Discovery Accelerator Grant to EP,
and JPL RSAs \# 1369565 and 1279141 to HS.
This work is based on observations made with the {\em Spitzer} Space Telescope, which is operated by the 
Jet Propulsion Laboratory, California Institute of Technology under a contract with NASA.
This research has made use of NASA's Astrophysics Data System.
The version of the ISO data presented in this paper correspond to the Highly Processed Data Product (HPDP) set called `Mid-IR Spectro Imaging ISOCAM CVF Observations'
by \citet{Boulanger_F_2005}, available for public use in the ISO Data Archive.

\bibliographystyle{mn2e}
\bibliography{reference}{}

\begin{thebibliography}{}
\makeatletter
\relax
\def\mn@urlcharsother{\let\do\@makeother \do\$\do\&\do\#\do\^\do\_\do\%\do\~}
\def\mn@doi{\begingroup\mn@urlcharsother \@ifnextchar[{\mn@doi@}{\mn@doi@[]}}
\def\mn@doi@[#1]#2{\def\@tempa{#1}\ifx\@tempa\@empty
  \href{http://dx.doi.org/#2}{doi:#2}\else \href{http://dx.doi.org/#2}{#1}\fi
  \endgroup}
\def\mn@eprint#1#2{\mn@eprint@#1:#2::\@nil}
\def\mn@eprint@arXiv#1{\href{http://arxiv.org/abs/#1}{{\tt arXiv:#1}}}
\def\mn@eprint@dblp#1{\href{http://dblp.uni-trier.de/rec/bibtex/#1.xml}{dblp:#1}}
\def\mn@eprint@#1:#2:#3:#4\@nil{\def\@tempa {#1}\def\@tempb {#2}\def\@tempc
  {#3}\ifx \@tempc \@empty \let\@tempc\@tempb \let\@tempb\@tempa \fi \ifx
  \@tempb \@empty \def\@tempb{arXiv}\fi \@ifundefined
  {mn@eprint@\@tempb}{\@tempb:\@tempc}{\expandafter \expandafter \csname
  mn@eprint@\@tempb\endcsname \expandafter{\@tempc}}}

\bibitem[\protect\citeauthoryear{{Allamandola}, {Tielens}  \&
  {Barker}}{{Allamandola} et~al.}{1989}]{Allamandola1989}
{Allamandola} L.~J.,  {Tielens} A.~G.~G.~M.,   {Barker} J.~R.,  1989, \mn@doi
  [\apjs] {10.1086/191396}, \href
  {http://adsabs.harvard.edu/abs/1989ApJS...71..733A} {71, 733}

\bibitem[\protect\citeauthoryear{{Bacon}, {Emsellem}, {Combes}, {Copin},
  {Monnet}  \& {Martin}}{{Bacon} et~al.}{2001}]{Bacon2001}
{Bacon} R.,  {Emsellem} E.,  {Combes} F.,  {Copin} Y.,  {Monnet} G.,   {Martin}
  P.,  2001, \mn@doi [\aap] {10.1051/0004-6361:20010317}, \href
  {http://adsabs.harvard.edu/abs/2001A%26A...371..409B} {371, 409}

\bibitem[\protect\citeauthoryear{{Barmby} et~al.,}{{Barmby}
  et~al.}{2006}]{Barmby2006lr}
{Barmby} P.,  et~al., 2006, \mn@doi [\apjl] {10.1086/508626}, \href
  {http://adsabs.harvard.edu/abs/2006ApJ...650L..45B} {650, L45}

\bibitem[\protect\citeauthoryear{{Beir{\~a}o}, {Brandl}, {Devost}, {Smith},
  {Hao}  \& {Houck}}{{Beir{\~a}o} et~al.}{2006}]{Beirao:06}
{Beir{\~a}o} P.,  {Brandl} B.~R.,  {Devost} D.,  {Smith} J.~D.,  {Hao} L.,
  {Houck} J.~R.,  2006, \mn@doi [\apjl] {10.1086/505027}, \href
  {http://adsabs.harvard.edu/abs/2006ApJ...643L...1B} {643, L1}

\bibitem[\protect\citeauthoryear{{Bender} et~al.,}{{Bender}
  et~al.}{2005}]{Bender2005}
{Bender} R.,  et~al., 2005, \mn@doi [\apj] {10.1086/432434}, \href
  {http://adsabs.harvard.edu/abs/2005ApJ...631..280B} {631, 280}

\bibitem[\protect\citeauthoryear{{Boulanger} et~al.,}{{Boulanger}
  et~al.}{2005}]{Boulanger_F_2005}
{Boulanger} F.,  et~al., 2005, \mn@doi [\aap] {10.1051/0004-6361:20047119},
  \href {http://adsabs.harvard.edu/abs/2005A%26A...436.1151B} {436, 1151}

\bibitem[\protect\citeauthoryear{{Brandl} et~al.,}{{Brandl}
  et~al.}{2006}]{Brandl2006}
{Brandl} B.~R.,  et~al., 2006, \mn@doi [\apj] {10.1086/508849}, \href
  {http://adsabs.harvard.edu/abs/2006ApJ...653.1129B} {653, 1129}

\bibitem[\protect\citeauthoryear{{Bregman}, {Bregman}  \& {Temi}}{{Bregman}
  et~al.}{2008}]{Bregman:08}
{Bregman} J.~D.,  {Bregman} J.~N.,   {Temi} P.,  2008, in {Chary} R.-R.,
  {Teplitz} H.~I.,   {Sheth} K.,  eds,  Astronomical Society of the Pacific
  Conference Series Vol. 381, Infrared Diagnostics of Galaxy Evolution. p.~34

\bibitem[\protect\citeauthoryear{{Calzetti} et~al.,}{{Calzetti}
  et~al.}{2010}]{Calzetti:2010fk}
{Calzetti} D.,  et~al., 2010, Conference Proceedings `Reionization to
  Exoplanets', ed. P. Ogle, ASP Conference Series, \href
  {http://adsabs.harvard.edu/abs/2010arXiv1005.4644C} {}

\bibitem[\protect\citeauthoryear{{Cesarsky} et~al.,}{{Cesarsky}
  et~al.}{1996}]{cesarsky1996}
{Cesarsky} C.~J.,  et~al., 1996, \aap, \href
  {http://adsabs.harvard.edu/abs/1996A%26A...315L..32C} {315, L32}

\bibitem[\protect\citeauthoryear{{Cesarsky}, {Lequeux}, {Pagani}, {Ryter},
  {Loinard}  \& {Sauvage}}{{Cesarsky} et~al.}{1998}]{1998Cesarsky}
{Cesarsky} D.,  {Lequeux} J.,  {Pagani} L.,  {Ryter} C.,  {Loinard} L.,
  {Sauvage} M.,  1998, \aap, \href
  {http://adsabs.harvard.edu/abs/1998A%26A...337L..35C} {337, L35}

\bibitem[\protect\citeauthoryear{{Croley}, {Barmby}, {Stock}, {Azimlu}  \&
  {Rosolowsky}}{{Croley} et~al.}{2015}]{Croley2015}
{Croley} M.,  {Barmby} P.,  {Stock} D.,  {Azimlu} M.,   {Rosolowsky} E.,  2015,
  \mnras, 0, submitted

\bibitem[\protect\citeauthoryear{{Davidge}, {Jensen}  \& {Olsen}}{{Davidge}
  et~al.}{2006}]{davidge06}
{Davidge} T.~J.,  {Jensen} J.~B.,   {Olsen} K.~A.~G.,  2006, \mn@doi [\aj]
  {10.1086/505463}, \href {http://adsabs.harvard.edu/abs/2006AJ....132..521D}
  {132, 521}

\bibitem[\protect\citeauthoryear{{Diamond-Stanic} \& {Rieke}}{{Diamond-Stanic}
  \& {Rieke}}{2010}]{Diamond2010}
{Diamond-Stanic} A.~M.,  {Rieke} G.~H.,  2010, \mn@doi [\apj]
  {10.1088/0004-637X/724/1/140}, \href
  {http://adsabs.harvard.edu/abs/2010ApJ...724..140D} {724, 140}

\bibitem[\protect\citeauthoryear{{Dong}, {Li}, {Wang}, {Lauer}, {Olsen},
  {Saha}, {Dalcanton}  \& {Williams}}{{Dong} et~al.}{2015}]{Dong:15}
{Dong} H.,  {Li} Z.,  {Wang} Q.~D.,  {Lauer} T.~R.,  {Olsen} K.~A.~G.,  {Saha}
  A.,  {Dalcanton} J.~J.,   {Williams} B.~F.,  2015, \mn@doi [\mnras]
  {10.1093/mnras/stv1256}, \href
  {http://adsabs.harvard.edu/abs/2015MNRAS.451.4126D} {451, 4126}

\bibitem[\protect\citeauthoryear{{Engelbracht}, {Rieke}, {Gordon}, {Smith},
  {Werner}, {Moustakas}, {Willmer}  \& {Vanzi}}{{Engelbracht}
  et~al.}{2008}]{Engelbracht_2008}
{Engelbracht} C.~W.,  {Rieke} G.~H.,  {Gordon} K.~D.,  {Smith} J.-D.~T.,
  {Werner} M.~W.,  {Moustakas} J.,  {Willmer} C.~N.~A.,   {Vanzi} L.,  2008,
  \mn@doi [\apj] {10.1086/529513}, \href
  {http://adsabs.harvard.edu/abs/2008ApJ...678..804E} {678, 804}

\bibitem[\protect\citeauthoryear{{Galliano}, {Madden}, {Tielens}, {Peeters}  \&
  {Jones}}{{Galliano} et~al.}{2008}]{Galliano2008}
{Galliano} F.,  {Madden} S.~C.,  {Tielens} A.~G.~G.~M.,  {Peeters} E.,
  {Jones} A.~P.,  2008, \mn@doi [\apj] {10.1086/587051}, \href
  {http://adsabs.harvard.edu/abs/2008ApJ...679..310G} {679, 310}

\bibitem[\protect\citeauthoryear{{Garcia} et~al.,}{{Garcia}
  et~al.}{2010}]{NucleusREF}
{Garcia} M.~R.,  et~al., 2010, \mn@doi [\apj] {10.1088/0004-637X/710/1/755},
  \href {http://adsabs.harvard.edu/abs/2010ApJ...710..755G} {710, 755}

\bibitem[\protect\citeauthoryear{{Geballe}, {Lacy}, {Persson}, {McGregor}  \&
  {Soifer}}{{Geballe} et~al.}{1985}]{Geballe:85}
{Geballe} T.~R.,  {Lacy} J.~H.,  {Persson} S.~E.,  {McGregor} P.~J.,   {Soifer}
  B.~T.,  1985, \apj, 292, 500

\bibitem[\protect\citeauthoryear{{Gezari}, {Pitts}  \& {Schmitz}}{{Gezari}
  et~al.}{1999}]{Gezari:99}
{Gezari} D.~Y.,  {Pitts} P.~S.,   {Schmitz} M.,  1999, VizieR Online Data
  Catalog, \href {http://adsabs.harvard.edu/abs/1999yCat.2225....0G} {2225, 0}

\bibitem[\protect\citeauthoryear{{Gillett}, {Forrest}  \& {Merrill}}{{Gillett}
  et~al.}{1973}]{Gillett:73}
{Gillett} F.~C.,  {Forrest} W.~J.,   {Merrill} K.~M.,  1973, \apj, 183, 87

\bibitem[\protect\citeauthoryear{{Gordon} et~al.,}{{Gordon}
  et~al.}{2006}]{gordon06a}
{Gordon} K.~D.,  et~al., 2006, \mn@doi [\apjl] {10.1086/501046}, \href
  {http://adsabs.harvard.edu/abs/2006ApJ...638L..87G} {638, L87}

\bibitem[\protect\citeauthoryear{{Gordon}, {Engelbracht}, {Rieke}, {Misselt},
  {Smith}  \& {Kennicutt}}{{Gordon} et~al.}{2008}]{Gordon:2008lr}
{Gordon} K.~D.,  {Engelbracht} C.~W.,  {Rieke} G.~H.,  {Misselt} K.~A.,
  {Smith} J.-D.~T.,   {Kennicutt} Jr. R.~C.,  2008, \mn@doi [\apj]
  {10.1086/589567}, \href {http://adsabs.harvard.edu/abs/2008ApJ...682..336G}
  {682, 336}

\bibitem[\protect\citeauthoryear{{Groves} et~al.,}{{Groves}
  et~al.}{2012}]{Groves2012}
{Groves} B.,  et~al., 2012, \mn@doi [\mnras]
  {10.1111/j.1365-2966.2012.21696.x}, \href
  {http://adsabs.harvard.edu/abs/2012MNRAS.426..892G} {426, 892}

\bibitem[\protect\citeauthoryear{{Hao} et~al.,}{{Hao} et~al.}{2005}]{Hao2005}
{Hao} L.,  et~al., 2005, \mn@doi [\apjl] {10.1086/431227}, \href
  {http://adsabs.harvard.edu/abs/2005ApJ...625L..75H} {625, L75}

\bibitem[\protect\citeauthoryear{{Hony}, {Van Kerckhoven}, {Peeters},
  {Tielens}, {Hudgins}  \& {Allamandola}}{{Hony} et~al.}{2001}]{Hony:oops:01}
{Hony} S.,  {Van Kerckhoven} C.,  {Peeters} E.,  {Tielens} A.~G.~G.~M.,
  {Hudgins} D.~M.,   {Allamandola} L.~J.,  2001, \aap, 370, 1030

\bibitem[\protect\citeauthoryear{{Houck} et~al.,}{{Houck}
  et~al.}{2004}]{IRS2004}
{Houck} J.~R.,  et~al., 2004, \mn@doi [\apjs] {10.1086/423134}, \href
  {http://adsabs.harvard.edu/abs/2004ApJS..154...18H} {154, 18}

\bibitem[\protect\citeauthoryear{{Hudgins} \& {Allamandola}}{{Hudgins} \&
  {Allamandola}}{2004}]{Hudgins:rev:04}
{Hudgins} D.~M.,  {Allamandola} L.~J.,  2004, in {Witt} A.~N.,  {Clayton}
  G.~C.,   {Draine} B.~T.,  eds,  Astronomical Society of the Pacific
  Conference Series Vol. 309, Astrophysics of Dust. p.~665

\bibitem[\protect\citeauthoryear{{Jarrett}, {Chester}, {Cutri}, {Schneider}  \&
  {Huchra}}{{Jarrett} et~al.}{2003}]{Jarrett:03}
{Jarrett} T.~H.,  {Chester} T.,  {Cutri} R.,  {Schneider} S.~E.,   {Huchra}
  J.~P.,  2003, \mn@doi [\aj] {10.1086/345794}, \href
  {http://adsabs.harvard.edu/abs/2003AJ....125..525J} {125, 525}

\bibitem[\protect\citeauthoryear{{Kaneda}, {Onaka}  \& {Sakon}}{{Kaneda}
  et~al.}{2005}]{Kaneda:05}
{Kaneda} H.,  {Onaka} T.,   {Sakon} I.,  2005, \mn@doi [\apjl]
  {10.1086/497913}, \href {http://adsabs.harvard.edu/abs/2005ApJ...632L..83K}
  {632, L83}

\bibitem[\protect\citeauthoryear{{Kaneda}, {Onaka}, {Sakon}, {Kitayama},
  {Okada}  \& {Suzuki}}{{Kaneda} et~al.}{2008}]{Kaneda:08}
{Kaneda} H.,  {Onaka} T.,  {Sakon} I.,  {Kitayama} T.,  {Okada} Y.,   {Suzuki}
  T.,  2008, \mn@doi [\apj] {10.1086/590243}, \href
  {http://adsabs.harvard.edu/abs/2008ApJ...684..270K} {684, 270}

\bibitem[\protect\citeauthoryear{{Kennicutt} Jr. et~al.,}{{Kennicutt}
  et~al.}{2003}]{kennicutt03}
{Kennicutt} Jr. R.~C.,  et~al., 2003, \mn@doi [\pasp] {10.1086/376941}, \href
  {http://adsabs.harvard.edu/abs/2003PASP..115..928K} {115, 928}

\bibitem[\protect\citeauthoryear{{Kessler} et~al.,}{{Kessler}
  et~al.}{1996}]{Kessler1996}
{Kessler} M.~F.,  et~al., 1996, \aap, \href
  {http://adsabs.harvard.edu/abs/1996A%26A...315L..27K} {315, L27}

\bibitem[\protect\citeauthoryear{{Lauer} et~al.,}{{Lauer}
  et~al.}{1993}]{Lauer1993}
{Lauer} T.~R.,  et~al., 1993, \mn@doi [\aj] {10.1086/116737}, \href
  {http://adsabs.harvard.edu/abs/1993AJ....106.1436L} {106, 1436}

\bibitem[\protect\citeauthoryear{{Li}, {Wang}  \& {Wakker}}{{Li}
  et~al.}{2009}]{Li09}
{Li} Z.,  {Wang} Q.~D.,   {Wakker} B.~P.,  2009, \mn@doi [\mnras]
  {10.1111/j.1365-2966.2009.14918.x}, \href
  {http://adsabs.harvard.edu/abs/2009MNRAS.397..148L} {397, 148}

\bibitem[\protect\citeauthoryear{{Li}, {Garcia}, {Forman}, {Jones}, {Kraft},
  {Lal}, {Murray}  \& {Wang}}{{Li} et~al.}{2011}]{Li2011}
{Li} Z.,  {Garcia} M.~R.,  {Forman} W.~R.,  {Jones} C.,  {Kraft} R.~P.,  {Lal}
  D.~V.,  {Murray} S.~S.,   {Wang} Q.~D.,  2011, \mn@doi [\apjl]
  {10.1088/2041-8205/728/1/L10}, \href
  {http://adsabs.harvard.edu/abs/2011ApJ...728L..10L} {728, L10}

\bibitem[\protect\citeauthoryear{{Madden}}{{Madden}}{2000}]{Madden:00}
{Madden} S.~C.,  2000, \mn@doi [\nar] {10.1016/S1387-6473(00)00050-6}, \href
  {http://adsabs.harvard.edu/abs/2000NewAR..44..249M} {44, 249}

\bibitem[\protect\citeauthoryear{{Mason}, {Levenson}, {Shi}, {Packham},
  {Gorjian}, {Cleary}, {Rhee}  \& {Werner}}{{Mason} et~al.}{2009}]{Mason2009}
{Mason} R.~E.,  {Levenson} N.~A.,  {Shi} Y.,  {Packham} C.,  {Gorjian} V.,
  {Cleary} K.,  {Rhee} J.,   {Werner} M.,  2009, \mn@doi [\apjl]
  {10.1088/0004-637X/693/2/L136}, \href
  {http://adsabs.harvard.edu/abs/2009ApJ...693L.136M} {693, L136}

\bibitem[\protect\citeauthoryear{{Mason} et~al.,}{{Mason}
  et~al.}{2012}]{Mason2012}
{Mason} R.~E.,  et~al., 2012, \mn@doi [\aj] {10.1088/0004-6256/144/1/11}, \href
  {http://adsabs.harvard.edu/abs/2012AJ....144...11M} {144, 11}

\bibitem[\protect\citeauthoryear{{McConnachie}, {Irwin}, {Ferguson}, {Ibata},
  {Lewis}  \& {Tanvir}}{{McConnachie} et~al.}{2005}]{Mcc2005}
{McConnachie} A.~W.,  {Irwin} M.~J.,  {Ferguson} A.~M.~N.,  {Ibata} R.~A.,
  {Lewis} G.~F.,   {Tanvir} N.,  2005, \mn@doi [\mnras]
  {10.1111/j.1365-2966.2004.08514.x}, \href
  {http://adsabs.harvard.edu/abs/2005MNRAS.356..979M} {356, 979}

\bibitem[\protect\citeauthoryear{{Melchior} \& {Combes}}{{Melchior} \&
  {Combes}}{2013}]{Melchior2013}
{Melchior} A.-L.,  {Combes} F.,  2013, \mn@doi [\aap]
  {10.1051/0004-6361/201220204}, \href
  {http://adsabs.harvard.edu/abs/2013A%26A...549A..27M} {549, A27}

\bibitem[\protect\citeauthoryear{{Moustakas}, {Kennicutt}, {Tremonti}, {Dale},
  {Smith}  \& {Calzetti}}{{Moustakas} et~al.}{2010}]{moustakas2010}
{Moustakas} J.,  {Kennicutt} Jr. R.~C.,  {Tremonti} C.~A.,  {Dale} D.~A.,
  {Smith} J.-D.~T.,   {Calzetti} D.,  2010, \mn@doi [\apjs]
  {10.1088/0067-0049/190/2/233}, \href
  {http://adsabs.harvard.edu/abs/2010ApJS..190..233M} {190, 233}

\bibitem[\protect\citeauthoryear{{Mu{\~n}oz-Mateos} et~al.,}{{Mu{\~n}oz-Mateos}
  et~al.}{2009}]{Munoz:09}
{Mu{\~n}oz-Mateos} J.~C.,  et~al., 2009, \mn@doi [\apj]
  {10.1088/0004-637X/703/2/1569}, \href
  {http://adsabs.harvard.edu/abs/2009ApJ...703.1569M} {703, 1569}

\bibitem[\protect\citeauthoryear{{Nagao}, {Maiolino}  \& {Marconi}}{{Nagao}
  et~al.}{2006}]{Nagao2006}
{Nagao} T.,  {Maiolino} R.,   {Marconi} A.,  2006, \mn@doi [\aap]
  {10.1051/0004-6361:20065216}, \href
  {http://adsabs.harvard.edu/abs/2006A%26A...459...85N} {459, 85}

\bibitem[\protect\citeauthoryear{{Pagani}, {Lequeux}, {Cesarsky}, {Milliard},
  {Lionard}  \& {Sauvage}}{{Pagani} et~al.}{1999}]{Pagani_1999}
{Pagani} L.,  {Lequeux} J.,  {Cesarsky} D.,  {Milliard} B.,  {Lionard} L.,
  {Sauvage} M.,  1999, \aap, 351, 447

\bibitem[\protect\citeauthoryear{{Peeters}}{{Peeters}}{2011}]{Peeters:toledo:11}
{Peeters} E.,  2011, in IAU Symposium. pp 149--161, \mn@eprint {arXiv}
  {1111.3680}, \mn@doi{10.1017/S174392131102494X}

\bibitem[\protect\citeauthoryear{{Puget} \& {Leger}}{{Puget} \&
  {Leger}}{1989}]{puget89}
{Puget} J.~L.,  {Leger} A.,  1989, \mn@doi [\araa]
  {10.1146/annurev.aa.27.090189.001113}, \href
  {http://adsabs.harvard.edu/abs/1989ARA%26A..27..161P} {27, 161}

\bibitem[\protect\citeauthoryear{{Roche}, {Aitken}, {Smith}  \& {Ward}}{{Roche}
  et~al.}{1991}]{Roche1991}
{Roche} P.~F.,  {Aitken} D.~K.,  {Smith} C.~H.,   {Ward} M.~J.,  1991, \mnras,
  \href {http://adsabs.harvard.edu/abs/1991MNRAS.248..606R} {248, 606}

\bibitem[\protect\citeauthoryear{{Saglia} et~al.,}{{Saglia}
  et~al.}{2010}]{Saglia:10}
{Saglia} R.~P.,  et~al., 2010, \mn@doi [\aap] {10.1051/0004-6361/200912805},
  \href {http://adsabs.harvard.edu/abs/2010A%26A...509A..61S} {509, A61}

\bibitem[\protect\citeauthoryear{{Sanders}, {Caldwell}, {McDowell}  \&
  {Harding}}{{Sanders} et~al.}{2012}]{Sanders_2011}
{Sanders} N.~E.,  {Caldwell} N.,  {McDowell} J.,   {Harding} P.,  2012, \mn@doi
  [\apj] {10.1088/0004-637X/758/2/133}, \href
  {http://adsabs.harvard.edu/abs/2012ApJ...758..133S} {758, 133}

\bibitem[\protect\citeauthoryear{{Sandstrom} et~al.,}{{Sandstrom}
  et~al.}{2012}]{Sandstrom12}
{Sandstrom} K.~M.,  et~al., 2012, \mn@doi [\apj] {10.1088/0004-637X/744/1/20},
  \href {http://adsabs.harvard.edu/abs/2012ApJ...744...20S} {744, 20}

\bibitem[\protect\citeauthoryear{{Sloan}, {Goes}, {Ramirez}, {Kraemer}  \&
  {Engelke}}{{Sloan} et~al.}{2015}]{Sloan:15}
{Sloan} G.~C.,  {Goes} C.~W.,  {Ramirez} R.~M.,  {Kraemer} K.~E.,   {Engelke}
  C.~W.,  2015, \apj, 0, submitted

\bibitem[\protect\citeauthoryear{{Smith} et~al.,}{{Smith}
  et~al.}{2007a}]{Smith:2007fk}
{Smith} J.-D.~T.,  et~al., 2007a, \mn@doi [\pasp] {10.1086/522634}, \href
  {http://adsabs.harvard.edu/abs/2007PASP..119.1133S} {119, 1133}

\bibitem[\protect\citeauthoryear{{Smith} et~al.,}{{Smith}
  et~al.}{2007b}]{Smith:2007lr}
{Smith} J.-D.~T.,  et~al., 2007b, \mn@doi [\apj] {10.1086/510549}, \href
  {http://adsabs.harvard.edu/abs/2007ApJ...656..770S} {656, 770}

\bibitem[\protect\citeauthoryear{{Smith} et~al.,}{{Smith}
  et~al.}{2010}]{Smith2010}
{Smith} H.~A.,  et~al., 2010, \mn@doi [\apj] {10.1088/0004-637X/716/1/490},
  \href {http://adsabs.harvard.edu/abs/2010ApJ...716..490S} {716, 490}

\bibitem[\protect\citeauthoryear{{Spitzer Science Center}}{{Spitzer Science
  Center}}{2012}]{SpitzerDAC}
{Spitzer Science Center} 2012, {Spitzer Data Analysis Cookbook}, v5.0.1 edn.
SSC, Pasadena, CA

\bibitem[\protect\citeauthoryear{{Spitzer Science Center}}{{Spitzer Science
  Center}}{2013}]{SpitzerIIH}
{Spitzer Science Center} 2013, {Spitzer-IRAC Instrument Handbook}, v2.0.3 edn.
SSC, Pasadena, CA

\bibitem[\protect\citeauthoryear{{Spoon}, {Marshall}, {Houck}, {Elitzur},
  {Hao}, {Armus}, {Brandl}  \& {Charmandaris}}{{Spoon}
  et~al.}{2007}]{Spoon2007}
{Spoon} H.~W.~W.,  {Marshall} J.~A.,  {Houck} J.~R.,  {Elitzur} M.,  {Hao} L.,
  {Armus} L.,  {Brandl} B.~R.,   {Charmandaris} V.,  2007, \mn@doi [\apjl]
  {10.1086/511268}, \href {http://adsabs.harvard.edu/abs/2007ApJ...654L..49S}
  {654, L49}

\bibitem[\protect\citeauthoryear{{Stock}, {Peeters}, {Tielens}, {Otaguro}  \&
  {Bik}}{{Stock} et~al.}{2013}]{Stock:13}
{Stock} D.~J.,  {Peeters} E.,  {Tielens} A.~G.~G.~M.,  {Otaguro} J.~N.,   {Bik}
  A.,  2013, \mn@doi [\apj] {10.1088/0004-637X/771/1/72}, \href
  {http://adsabs.harvard.edu/abs/2013ApJ...771...72S} {771, 72}

\bibitem[\protect\citeauthoryear{{Sturm} et~al.,}{{Sturm}
  et~al.}{2005}]{Sturm2005}
{Sturm} E.,  et~al., 2005, \mn@doi [\apjl] {10.1086/444359}, \href
  {http://adsabs.harvard.edu/abs/2005ApJ...629L..21S} {629, L21}

\bibitem[\protect\citeauthoryear{{Tielens}}{{Tielens}}{2008}]{Tielens2008}
{Tielens} A.~G.~G.~M.,  2008, \mn@doi [\araa]
  {10.1146/annurev.astro.46.060407.145211}, \href
  {http://adsabs.harvard.edu/abs/2008ARA%26A..46..289T} {46, 289}

\bibitem[\protect\citeauthoryear{{Vega} et~al.,}{{Vega} et~al.}{2010}]{Vega:10}
{Vega} O.,  et~al., 2010, \mn@doi [\apj] {10.1088/0004-637X/721/2/1090}, \href
  {http://adsabs.harvard.edu/abs/2010ApJ...721.1090V} {721, 1090}

\bibitem[\protect\citeauthoryear{{Vermeij}, {Peeters}, {Tielens}  \& {van der
  Hulst}}{{Vermeij} et~al.}{2002}]{Vermeij2002}
{Vermeij} R.,  {Peeters} E.,  {Tielens} A.~G.~G.~M.,   {van der Hulst} J.~M.,
  2002, \mn@doi [\aap] {10.1051/0004-6361:20011628}, \href
  {http://adsabs.harvard.edu/abs/2002A%26A...382.1042V} {382, 1042}

\bibitem[\protect\citeauthoryear{{Villaume}, {Conroy}  \& {Johnson}}{{Villaume}
  et~al.}{2015}]{Villaume:15}
{Villaume} A.,  {Conroy} C.,   {Johnson} B.~D.,  2015, \mn@doi [\apj]
  {10.1088/0004-637X/806/1/82}, \href
  {http://adsabs.harvard.edu/abs/2015ApJ...806...82V} {806, 82}

\bibitem[\protect\citeauthoryear{{Werner} et~al.,}{{Werner}
  et~al.}{2004}]{spitzer2004}
{Werner} M.~W.,  et~al., 2004, \mn@doi [\apjs] {10.1086/422992}, \href
  {http://adsabs.harvard.edu/abs/2004ApJS..154....1W} {154, 1}

\makeatother
\end{thebibliography}

\bsp

\label{lastpage}

\end{document}